\documentclass[a4paper]{jpconf}

\usepackage{graphicx}
\usepackage{amsmath,amssymb}
\usepackage{subfigure}
\usepackage{color}
\usepackage{multirow}
\usepackage{cite}
\usepackage{bbold}
\usepackage{bm}

\begin{document}

\title{Low-energy parameters and spin gap of a frustrated spin-$s$ Heisenberg antiferromagnet with $s \leq \frac{3}{2}$ on the honeycomb lattice}


\author{\underline{R F Bishop} and P H Y Li} 

\address{School of Physics and Astronomy, Schuster Building, The University of Manchester, Manchester, M13 9PL, UK}
\address{School of Physics and Astronomy, University of Minnesota, 116 Church Street SE, Minneapolis, Minnesota 55455, USA}

\ead{raymond.bishop@manchester.ac.uk; peggyhyli@gmail.com}

\begin{abstract}
The coupled cluster method is implemented at high orders of approximation to investigate the zero-temperature $(T=0)$ phase diagram of the frustrated spin-$s$ $J_{1}$--$J_{2}$--$J_{3}$ antiferromagnet on the honeycomb lattice.  The system has isotropic Heisenberg interactions of strength $J_{1}>0$, $J_{2}>0$ and $J_{3}>0$ between nearest-neighbour, next-nearest-neighbour and next-next-nearest-neighbour pairs of spins, respectively.  We study it in the case $J_{3}=J_{2}\equiv \kappa J_{1}$, in the window $0 \leq \kappa \leq 1$ that contains the classical tricritical point (at $\kappa_{{\rm cl}}=\frac{1}{2}$) of maximal frustration, appropriate to the limiting value $s \to \infty$ of the spin quantum number.  We present results for the magnetic order parameter $M$, the triplet spin gap $\Delta$, the spin stiffness $\rho_{s}$ and the zero-field transverse magnetic susceptibility $\chi$ for the two collinear quasiclassical antiferromagnetic (AFM) phases with N\'{e}el and striped order, respectively.  Results for $M$ and $\Delta$ are given for the three cases $s=\frac{1}{2}$, $s=1$ and $s=\frac{3}{2}$, while those for $\rho_{s}$ and $\chi$ are given for the two cases $s=\frac{1}{2}$ and $s=1$.  On the basis of all these results we find that the spin-$\frac{1}{2}$ and spin-1 models both have an intermediate paramagnetic phase, with no discernible magnetic long-range order, between the two AFM phases in their $T=0$ phase diagrams, while for $s > 1$ there is a direct transition between them.  Accurate values are found for all of the associated quantum critical points.  While the results also provide strong evidence for the intermediate phase being gapped for the case $s=\frac{1}{2}$, they are less conclusive for the case $s=1$.  On balance however, at least the transition in the latter case at the striped phase boundary seems to be to a gapped intermediate state.
\end{abstract}

\section{Introduction}
\label{intro_sec}

Extended, uniform spin-lattice models of quantum magnets comprise a
number $N (\to \infty)$ of SU(2) spins with a given spin quantum
number $s$ placed on the sites of a specified regular, periodic
lattice in $d$ dimensions.  The interactions between the spins are
typically modelled by pairwise Heisenberg exchange interactions with
specified coupling strengths.  Of particular interest in this context
are models that exhibit both quasiclassical behaviour and nonclassical
behaviour, typically with paramagnetic phases in some part of the
relevant parameter space that do not exist for their classical
counterparts (i.e., with $s \to \infty$), and which hence do not show
magnetic long-range order (LRO).  

Clearly, it is therefore especially advantageous to investigate models
and situations in which quantum effects are enhanced.  Broadly
speaking, one expects that quantum fluctuations will be greater in
such spin-lattice systems where each of the spin quantum number $s$,
the dimensionality $d$, and the coordination number $z$ of the lattice
take lower values.  However, the well-known Mermin-Wagner theorem
\cite{Mermin:1966} prohibits all forms of magnetic LRO for systems
with $d=1$, even at zero temperature ($T=0$), and for systems with
$d=2$ and $T\neq 0$, since for all such systems it is not possible to
break a continuous symmetry, and all such quasiclassical states with
magnetic LRO break both time-reversal symmetry and SU(2) spin-rotation
symmetry.  Hence, two-dimensional (2D) models at $T=0$ have come to
occupy a key role in the study of quantum phase transitions (QPTs)
\cite{Sachdev:2008_QPT,Sachdev:2011_QPT}.

Since the honeycomb lattice has the lowest coordination number
($z = 3$) of all eleven 2D Archimedean lattices (i.e., those that comprise
only regular polygons, possibly of different sorts, and with all
sites equivalent to one another), spin-lattice models based on it have
been the subject of intense interest in recent years.  Prototypical
such models that have been investigated are those in which the
Hamiltonian includes only terms in which the spins at lattice sites
$i$ and $j$ interact via an isotropic Heisenberg interaction of the
form $J_{ij}\mathbf{s}_{i}\cdot\mathbf{s}_{j}$.  Much attention has
been focussed on the case when the exchange couplings $J_{ij}$ are
restricted to be between nearest-neighbour (NN) pairs, all with equal
strength $J_{1}$, next-nearest-neighbour (NNN) pairs, all with equal
strength $J_{2}$, and next-next-nearest-neighbour (NNNN) pairs, all
with equal strength $J_{3}$.  Both the resulting so-called
$J_{1}$--$J_{2}$--$J_{3}$ model and, particularly, the two special
cases of it with $J_{3}=0$ and $J_{3}=J_{2}$ have been extensively
investigated with a wide variety of theoretical techniques.  While
frustrated such honeycomb lattice models with spins having
$s=\frac{1}{2}$ have been particularly intensively studied
\cite{Rastelli:1979_honey,Mattsson:1994_honey,Fouet:2001_honey,Mulder:2010_honey,Wang:2010_honey,Cabra:2011_honey,Ganesh:2011_honey_merge,Ganesh:2011_honey_errata_merge,Clark:2011_honey,DJJF:2011_honeycomb,Reuther:2011_honey,Albuquerque:2011_honey,Mosadeq:2011_honey,Oitmaa:2011_honey,Mezzacapo:2012_honey,PHYLi:2012_honeycomb_J1neg,Bishop:2012_honeyJ1-J2,Bishop:2012_honey_circle-phase,Li:2012_honey_full,RFB:2013_hcomb_SDVBC,Ganesh:2013_honey_J1J2mod-XXX,Zhu:2013_honey_J1J2mod-XXZ,Zhang:2013_honey,Gong:2013_J1J2mod-XXX,Yu:2014_honey_J1J2mod,Bishop:2015_honey_low-E-param},
considerably less attention has been paid to their counterparts with
$s > \frac{1}{2}$
\cite{Zhao:2012_honeycomb_s1,Gong:2015_honey_J1J2mod_s1,Bishop:2016_honey_grtSpins,Li:2016_honey_grtSpins,Li:2016_honeyJ1-J2_s1}.

Our main aim in the present work is to extend the study of such
honeycomb-lattice models with $s \geq 1$, by applying to them the
coupled cluster method (CCM) (see, e.g., Refs.\
\cite{Bishop:1987_ccm,Bishop:1991_TheorChimActa_QMBT,Bishop:1998_QMBT_coll,Bartlett:2007_ccm})
implemented to high orders of approximation.  The CCM has been shown
to give results of unsurpassed accuracy to an extremely wide array of
physical systems, including those in condensed matter physics, quantum
chemistry, atomic and molecular physics, quantum optics and
solid-state optoelectronics, and nuclear and subnuclear physics (and see,
e.g., Refs.\
\cite{Bishop:1987_ccm,Bishop:1991_TheorChimActa_QMBT,Bishop:1998_QMBT_coll,Bartlett:2007_ccm,Coester:1958_ccm,Coester:1960_ccm,Cizek:1966_ccm,Cizek:1969_ccm,Bishop:1978_ccm,Kummel:1978_ccm,Bishop:1982_ccm,Arponen:1983_ccm,Arponen:1987_ccm,Arponen:1987_ccm_2,Arponen:1991_ccm,Stanton:1993_ccm,Arponen:1993_ccm,Arponen:1993_ccm_b,Zeng:1998_SqLatt_TrianLatt,Fa:2004_QM-coll,Bishop:2014_honey_XXZ_nmp14,Bishop:2017_SqLatt_XXZ}
and references cited therein).  We note in particular that the CCM has
already been applied with great success to a wide diversity of
spin-lattice problems of interest in quantum magnetism (and see, e.g.,
Refs.\
\cite{DJJF:2011_honeycomb,PHYLi:2012_honeycomb_J1neg,Bishop:2012_honeyJ1-J2,Bishop:2012_honey_circle-phase,Li:2012_honey_full,RFB:2013_hcomb_SDVBC,Bishop:2015_honey_low-E-param,Bishop:2016_honey_grtSpins,Li:2016_honey_grtSpins,Li:2016_honeyJ1-J2_s1,Zeng:1998_SqLatt_TrianLatt,Fa:2004_QM-coll,Bishop:2014_honey_XXZ_nmp14,Bishop:2017_SqLatt_XXZ}
and references cited therein).  These include specific applications \cite{DJJF:2011_honeycomb,Bishop:2012_honey_circle-phase,Bishop:2015_honey_low-E-param,Li:2016_honey_grtSpins} to the model that we study further here, as discussed in more detail in Sec.\ \ref{model_sec}.

The plan for the reminder of this paper is as follows.  In Sec.\
\ref{model_sec} we first describe the model, including some of the
pertinent results for both the classical ($s \to \infty$) and
$s=\frac{1}{2}$ cases.  The CCM is then briefly reviewed in Sec.\
\ref{ccm_section}, before we present in Sec.\ \ref{results_sec} our
results using it for the particular cases $s=1$ and $s=\frac{3}{2}$.
In particular, we present results for the ground-state (GS) magnetic
order parameter, the triplet spin gap, the GS spin stiffness and the
GS zero-field transverse magnetic susceptibility.  Finally, we
conclude with a discussion of the results in Sec.\
\ref{conclusions_sec}.

\section{The model}
\label{model_sec}
\begin{figure*}[!t]
\hspace{1.5cm}
\mbox{
\subfigure[]{\includegraphics[width=4.5cm]{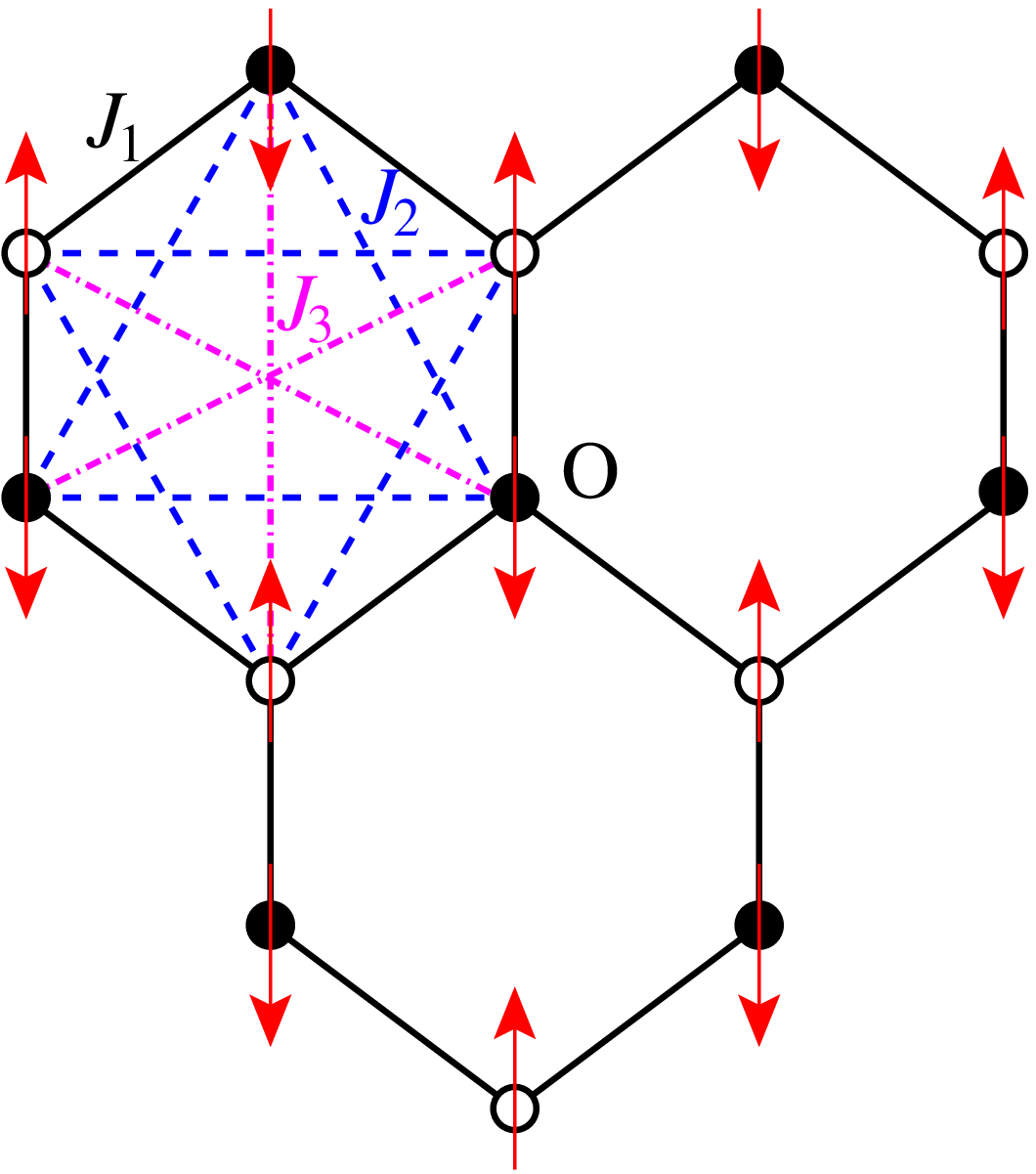}}
\quad \quad \subfigure[]{\includegraphics[width=4.5cm]{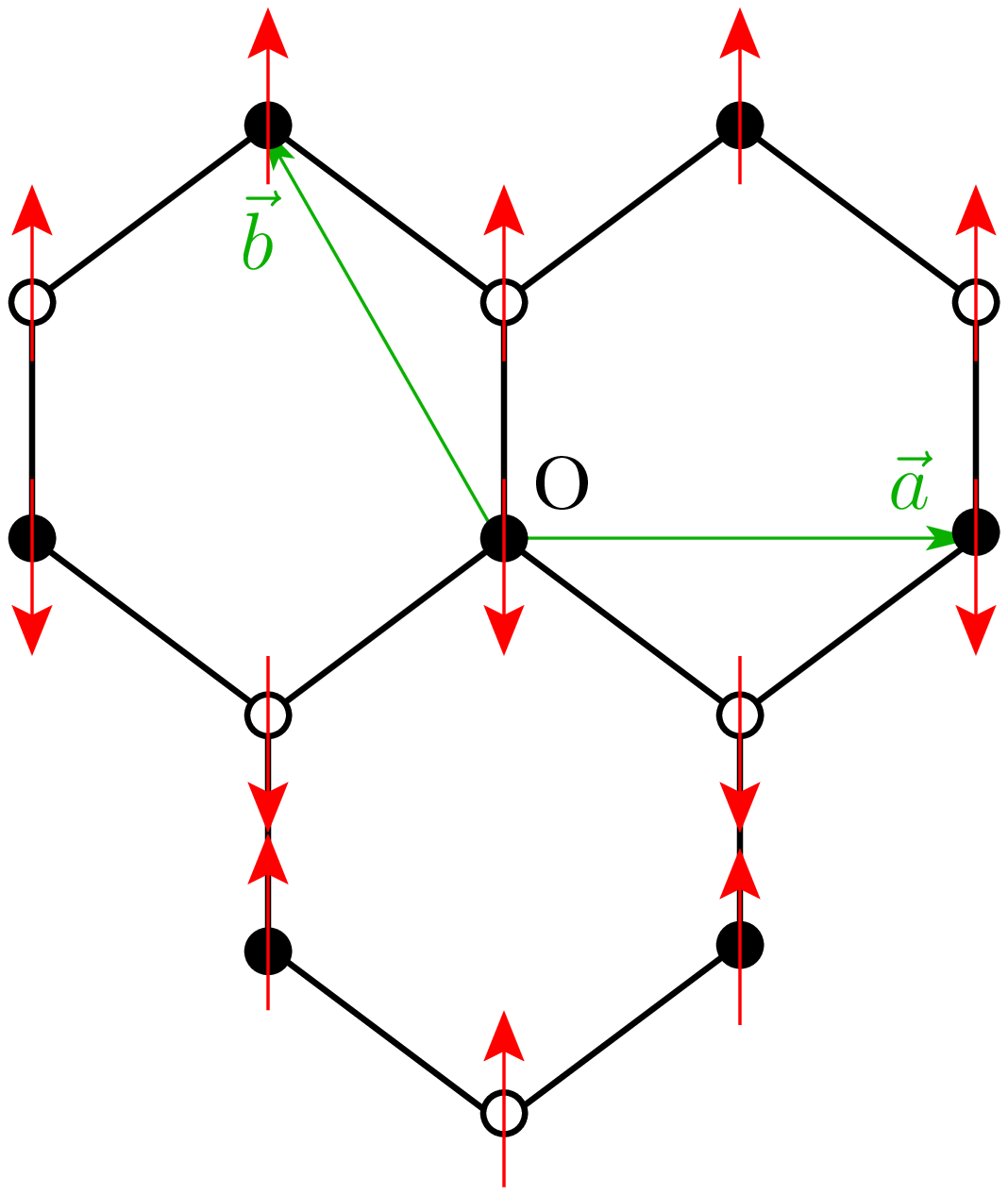}}
\quad \quad \subfigure{\includegraphics[width=3cm]{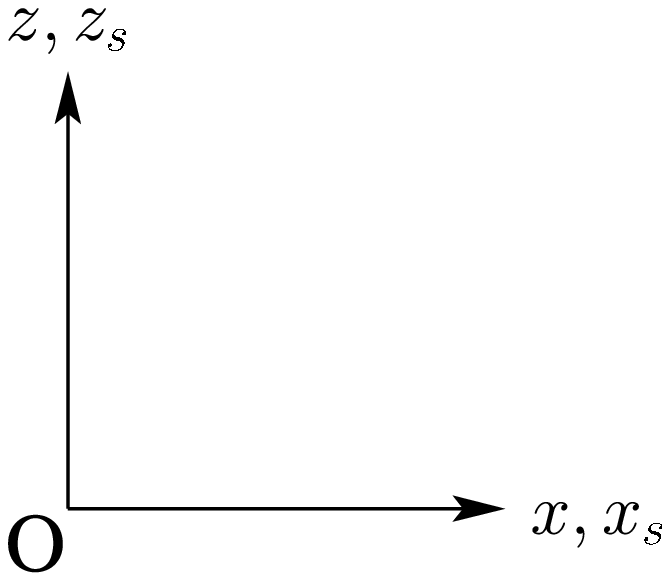}}
}
\caption{The $J_{1}$--$J_{2}$--$J_{3}$ honeycomb model
  with $J_{1}>0; J_{2}>0; J_{3}>0$, showing (a) the bonds ($J_{1} =$ 
  -----~; $J_{2} = - - -$~; $J_{3} = - \cdot -$) and the
  N\'{e}el state, and (b) triangular Bravais lattice vectors
  $\mathbf{a}$ and $\mathbf{b}$ and one of the three equivalent
  striped states.  Sites on sublattices ${\cal A}$ and ${\cal B}$ are
  shown by filled and empty circles respectively, and spins on
  the lattice are represented by the (red) arrows on the lattice sites.  We also show both the lattice-plane axes $(x,z)$ and the spin-space axes $(x_{s},z_{s})$.}
\label{model_pattern}
\end{figure*}

The Hamiltonian of the $J_{1}$--$J_{2}$--$J_{3}$ model on the
honeycomb lattice is specified as
\begin{equation}
\begin{aligned}
H & =  J_{1}{\displaystyle\sum_{{\langle i,j \rangle}}} \mathbf{s}_{i}\cdot\mathbf{s}_{j} + 
J_{2}{\displaystyle \sum_{{\langle\langle i,k \rangle\rangle}}} \mathbf{s}_{i}\cdot\mathbf{s}_{k} + 
J_{3}{\displaystyle \sum_{{\langle\langle\langle i,l \rangle\rangle\rangle}}} \mathbf{s}_{i}\cdot\mathbf{s}_{l}  \\
& \equiv  J_{1}h(x,y)\,; \quad  x \equiv J_{2}/J_{1}\,, \quad y \equiv J_{3}/J_{1}\,,
\label{H_eq}
\end{aligned}
\end{equation}
where the operators
${\bf s}_{i} \equiv (s^{x}_{i}, s^{y}_{i}, s^{z}_{i})$ are the usual
SU(2) quantum spin operators on lattice site $i$, with
${\bf s}^{2}_{i} = s(s+1)\mathbb{1}$.  We shall compare and contrast
here the cases with $s=\frac{1}{2}$, $s=1$ and $s\geq\frac{3}{2}$.
The sums in Eq.\ (\ref{H_eq}) over $\langle i,j \rangle$,
$\langle \langle i,k \rangle \rangle$ and
$\langle \langle \langle i,l \rangle \rangle \rangle$ run over all
NN, NNN and NNNN bonds respectively, counting each bond once only in
each sum.  The lattice and the exchange bonds are illustrated in Fig.\
\ref{model_pattern}(a).
We shall be interested in the present paper in the case when all three
bonds are antiferromagnetic (AFM) (i.e., when $J_{i}>0$; $i=1,2,3$).
The NN exchange coupling constant $J_{1}$ simply sets the overall
energy scale, and hence the nontrivial parameters of the model may be
taken as $J_{2}/J_{1} \equiv x$ and $J_{3}/J_{1} \equiv y$, so that we
may rewrite the Hamiltonian as $H \equiv J_{1}h(x,y)$, as in Eq.\
(\ref{H_eq}).

As we discuss in more detail below, the classical version ($s \to \infty$) of
the model has a tricritical point at $x=y=\frac{1}{2}$ in its $T=0$ phase
diagram \cite{Rastelli:1979_honey,Fouet:2001_honey}).  At this point
the Hamiltonian may be rewritten in the particularly simple form
\begin{equation}
h(\tfrac{1}{2},\tfrac{1}{2}) = \frac{1}{4}\sum_{\chi}\mathbf{s}^{2}_{\chi}-\frac{3}{4}Ns(s+1)\,,
\end{equation}
of a sum over edge-sharing hexagonal plaquettes $\chi$, up to an additive constant term, where
\begin{equation}
\mathbf{s}_{\chi}=\sum_{i \in \chi}\mathbf{s}_{i}
\end{equation}
is the total spin of the six spins on the elementary hexagon $\chi$.
Clearly, at this special point $x=y=\frac{1}{2}$, any state wth zero
total spin on each elementary hexagonal plaquette is a classical $T=0$
GS phase, thereby giving rise to a macroscopic degeneracy.  In fact,
as we describe below, one of the three classical states with perfect
magnetic LRO that emerges from this tricritical point, is itself also
associated with an infinitely degenerate family (IDF) of GS phases.
Indeed, this is one of the main reasons to be interested in the model,
since classical spin-lattice systems that exhibit such an IDF of $T=0$
GS phases are, {\it a priori}, prime candidates for the possible
emergence of novel quantum phases (i.e., without classical
counterparts) in the cases when the spin quantum number $s$ is finite,
since the role of quantum fluctuations is then, self-evidently,
enhanced.

While the honeycomb lattice is bipartite (and hence not geometrically
frustrated) it is non-Bravais.  It comprises two sites per unit cell,
with two triangular Bravais sublattices $\mathcal{A}$ and
$\mathcal{B}$.  If we define the lattice to occupy the $xz$ plane, the
basis vectors may be chosen as $\mathbf{a}=\sqrt{3}d\hat{x}$ and
$\mathbf{b}=\frac{1}{2}(-\sqrt{3}\hat{x}+3\hat{z})d$, as illustrated
in Fig.\ \ref{model_pattern}(b), where $\hat{x}$ and $\hat{z}$ are
unit vectors in the $x$ and $z$ directions respectively, and $d$ is
the honeycomb lattice spacing (i.e., the distance between NN sites).
The unit cell $i$ residing at position vector
$\mathbf{R}_{i} \equiv m_{i}\mathbf{a} + n_{i}\mathbf{b}$, with
$m_{i},n_{i}\in\mathbb{Z}$, contains the two sites at
$\mathbf{R}_{i} \in \mathcal{A}$ and
$(\mathbf{R}_{i}+d\hat{z})\in\mathcal{B}$.  The corresponding
reciprocal lattice vectors are hence given by
$\boldsymbol{\alpha}=2\pi(\sqrt{3}\hat{x}+\hat{z})/(3d)$ and
$\boldsymbol{\beta}=4\pi/(3d)\hat{z}$.  Thus, the first Brillouin zone
and the Wigner-Seitz unit cell are the respective parallelograms
formed by the pairs of vectors
$(\boldsymbol{\alpha},\boldsymbol{\beta})$ and
$(\mathbf{a},\mathbf{b})$.  Clearly, one may also take both as being
centred on a point of sixfold rotational symmetry in their
corresponding lattice planes.  Thus, just as the Wigner-Seitz unit
cell may be chosen to be bounded by the sides of an elementary hexagon
(i.e., of side length $d$), as in Fig.\ \ref{model_pattern}, so the
first Brillouin zone may also be chosen as a hexagon of side length
$4\pi/(3\sqrt{3}d)$, which is rotated by $90^{\circ}$ with respect to
the Wigner-Seitz hexagon.  Thus, the corners of the hexagon forming
the first Brillouin zone are given by the vectors
\begin{equation}
\mathbf{K}^{(1)}=\frac{4\pi}{3\sqrt{3}d}\hat{x}\,, \quad \mathbf{K}^{(2)}=\frac{2\pi}{3\sqrt{3}d}(\hat{x}+\sqrt{3}\hat{z})\,, \quad \mathbf{K}^{(3)}=\frac{2\pi}{3\sqrt{3}d}(-\hat{x}+\sqrt{3}\hat{z})\,,
\end{equation}
with the remaining three corners at positions
$\mathbf{K}^{(i+3)}=-\mathbf{K}^{(i)}$; $i=1,2,3$.  Similarly, the
mid-points of the edges of the first Brillouin zone have the vectors
\begin{equation}
\mathbf{M}^{(1)}=\frac{\pi}{3d}(\sqrt{3}\hat{x}+\hat{z})\,, \quad \mathbf{M}^{(2)}=\frac{2\pi}{3d}\hat{z}\,, \quad \mathbf{M}^{(3)}=\frac{\pi}{3d}(-\sqrt{3}\hat{x}+\hat{z})\,,  \label{eq_midPoints_firstBrillouinZone}
\end{equation}      
with the remaining three midpoints at positions
$\mathbf{M}^{(i+3)}=-\mathbf{M}^{(i)}$; $i=1,2,3$.

Returning now to the classical $(s \to \infty)$ version of the model
of Eq.\ (\ref{H_eq}), it has been shown
\cite{Rastelli:1979_honey,Fouet:2001_honey} that its generic stable GS
phase is described by a coplanar spiral configuration of spins.  This
may be defined, as usual, by a wave vector $\mathbf{Q}$, together with
(for the non-Bravais two-site unit cell of the model) an angle $\phi$
that defines the relative orientation of the two sites in the same
unit cell at position vector $\mathbf{R}_{i}$.  The two classical
spins in unit cell $i$ thus have orientations
\begin{equation}
\mathbf{s}_{i,\sigma}=-s[\cos(\mathbf{Q}\cdot\mathbf{R}_{i}+\phi_{\sigma})\hat{z}_{s}+\sin(\mathbf{Q}\cdot\mathbf{R}_{i}+\phi_{\sigma})\hat{x}_{s}]\,; \quad \sigma=\mathcal{A},\mathcal{B}\,,  \label{eq_classical_spins_orient}
\end{equation}
where the spin-space plane is now defined by the two orthogonal unit
vectors $\hat{x}_{s}$ and $\hat{z}_{s}$, shown in Fig.\
\ref{model_pattern}.  The two angles $\phi_{\sigma}$ are chosen so that
$\phi_{\mathcal{A}}=0$ and $\phi_{\mathcal{B}}=\phi$.

In the case considered here when all three bonds are AFM (i.e., with $J_{1}>0$, $x>0$, $y>0$), it has been shown \cite{Rastelli:1979_honey,Fouet:2001_honey} that the $T=0$ classical GS phase diagram comprises three phases that meet at the tricritical point $(x,y) = (\frac{1}{2},\frac{1}{2})$, discussed above.  If we define our lattice origin to be at the centre of the hexagonal Wigner-Seitz unit cell, one may show that one value of the spiral wave vector $\mathbf{Q}$ that minimizes the classical GS energy in this case is
\begin{equation}
\mathbf{Q} = \frac{2}{\sqrt{3}d}\cos^{-1}\left[\frac{(1-2x)}{4(x-y)}\right]\hat{x}\,,  \label{eq_spiral_wave_vect}
\end{equation}
together with the value $\phi=\pi$.  Clearly, for Eq.\ (\ref{eq_spiral_wave_vect}) to be physically valid, we require $|(1-2x)/(x-y)|\leq 4$, or equivalently,
\begin{equation}
y \leq \frac{3}{2}x - \frac{1}{4}\,; \quad y \leq \frac{1}{2}x + \frac{1}{4}\,.  \label{eq_spiral_wave_vect_validCondition}
\end{equation}

The two straight lines defined by Eq.\
(\ref{eq_spiral_wave_vect_validCondition}) taken as equalities cross
at the (tricritical) point $(x,y)=(\frac{1}{2},\frac{1}{2})$.  Along
the first boundary line, $y=\frac{3}{2}x-\frac{1}{4}$,
$\mathbf{Q}=\mathbf{\Gamma}=(0,0)$, which (again, together with the
minimizing condition $\phi=\pi$) simply describes the N\'{e}el phase
illustrated in Fig.\ \ref{model_pattern}(a).  Similarly, everywhere
along the second boundary line in Eq.\
(\ref{eq_spiral_wave_vect_validCondition}),
$y=\frac{1}{2}x + \frac{1}{4}$, we have
$\mathbf{Q}=2\pi/(\sqrt{3}d)\hat{x}$.  In fact this vector lies
outside the first Brillouin zone.  When mapped back inside it takes
the equivalent value $\mathbf{Q}=\mathbf{M}^{(2)}$, one of the
midpoints of the edges of the hexagonal first Brillouin zone, given in
Eq.\ (\ref{eq_midPoints_firstBrillouinZone}).  This wave vector (again
together with the minimizing condition $\phi=0$) describes the striped
AFM phase illustrated in Fig.\ \ref{model_pattern}(b).  It is clear
that the phase transitions across the boundaries of Eq.\
(\ref{eq_spiral_wave_vect_validCondition}) considered as equalities
(i.e., between the N\'{e}el and spiral phases, and between the striped
and spiral phases) are both continuous ones.  One may also readily
show from energetics that the phase transition between the two
collinear AFM (N\'{e}el and striped) phases is of first-order type, and
occurs along the boundary $x=\frac{1}{2}$, which again meets the other
two phase boundaries at the classical tricritical point
$(x,y)=(\frac{1}{2},\frac{1}{2})$.

It is obvious that both the striped phase and the spiral phase
described by Eq.\ (\ref{eq_spiral_wave_vect}) break the rotational
symmetry and that there must be two other equivalent states in each
case, obtained by a rotation of $\pm\frac{2}{3}\pi$ in the honeycomb
$xz$ plane.  For the striped state this threefold degeneracy is simply
equivalent to the wave vector $\mathbf{Q}$ being allowed to take any of the
position vectors $\mathbf{M}^{(i)}$; $i=1,2,3$, of the three inequivalent
midpoints of the sides of the hexagonal first Brillouin zone, as given
by Eq.\ (\ref{eq_midPoints_firstBrillouinZone}), together with
$\phi=0$ in the cases with $i=1,3$.

Although it is known \cite{Villain:1977_ordByDisord} that classical GS
spin configurations can generally be described as in Eq.\
(\ref{eq_classical_spins_orient}), it is also known
\cite{Villain:1977_ordByDisord} that there are exceptional cases when
the GS order is either not unique (up to a global rotation) or has a
discrete degeneracy such as that associated with the striped state.
Such exceptions are known
\cite{Fouet:2001_honey,Villain:1977_ordByDisord} to occur for special
values of the ordering wave vector $\mathbf{Q}$.  These include the
cases when $\mathbf{Q}$ takes a value equal to either one half or one
quarter of a reciprocal lattice vector
$\mathbf{G}_{i}\equiv k_{i}\boldsymbol{\alpha} +
l_{i}\boldsymbol{\beta}$, with $k_{i},l_{i} \in \mathbb{Z}$.  The
striped states are precisely of this form since their wave vectors
$\mathbf{Q}=\mathbf{M}^{(i)}$, $i=1,2,3$, are precisely one half of
corresponding reciprocal lattice vectors, i.e.,
$\mathbf{M}^{(1)}=\frac{1}{2}\boldsymbol{\alpha}$,
$\mathbf{M}^{(2)}=\frac{1}{2}\boldsymbol{\beta}$,
$\mathbf{M}^{(3)}=\frac{1}{2}(\boldsymbol{\beta}-\boldsymbol{\alpha})$.
In this case it has been explicitly shown \cite{Fouet:2001_honey} that the
GS ordering has an IDF of non-planar spin configurations, all
degenerate in energy with the collinear striped states.  It has also
been shown \cite{Rastelli:1979_honey,Fouet:2001_honey} that, at least
in the large-$s$ limit when lowest-order spin-wave theory (LSWT)
becomes exact, quantum fluctuations lift this degeneracy in favour of
the AFM striped states, which now have the lowest energy.

To conclude our discussion of the classical limit $(s \to \infty)$ of
the $J_{1}$--$J_{2}$--$J_{3}$ model on the honeycomb lattice in the
region where $J_{1}>0$, $x \geq 0$, $y \geq 0$, we have seen that at
$T=0$ its classical GS phase diagram has three phases, each with
perfect magnetic LRO.  These comprise (a) a N\'{e}el AFM phase in the
region $y>0$, $0 < x < \frac{1}{6}$ and
$y > \frac{3}{2}x - \frac{1}{4}$, $\frac{1}{6} < x < \frac{1}{2}$; (b)
a collinear striped AFM phase in the region
$y > \frac{1}{2}x + \frac{1}{4}$, $x > \frac{1}{2}$; and (c) a
coplanar spiral phase in the region
$0 < y < \frac{3}{2}x - \frac{1}{4}$, $\frac{1}{6} < x < \frac{1}{2}$
and $0 < y < \frac{1}{2}x + \frac{1}{4}$, $x > \frac{1}{2}$.  For a
further discussion of the classical $T=0$ GS phase diagram of the
model, and for other exceptional cases (including those when
$\mathbf{Q}$ takes a value $\frac{1}{2}\mathbf{G}_{i}$ or
$\frac{1}{4}\mathbf{G}_{i})$ the reader is referred to Refs.\
\cite{Fouet:2001_honey,Mulder:2010_honey,Li:2016_honey_grtSpins}.
Although not of direct relevance here, the case $y=0$ that includes a
spiral phase for $x>\frac{1}{6}$, also includes a one-parameter IDF of
incommensurate GS phases wherein the wave vector $\mathbf{Q}$ can
orient in an arbitrary direction, with degenerate solutions along a
specific contour for a given value of $x$ (and see Refs.\
\cite{Fouet:2001_honey,Mulder:2010_honey} for details).  Again, at the
level of LSWT, this degeneracy is lifted by quantum fluctuations to
give spiral order by disorder.  

The most interesting region of the classical phase diagram, for
reasons discussed above, includes the tricritical point and the
striped phase (that is part of an IDF of GS phases).  Both are sampled
along the line $y=x=\kappa$ (i.e., in the region $J_{1}>0$,
$J_{3}=J_{2}\equiv \kappa J_{1}$, $\kappa > 0$).  Hence, for
the remainder of this study we work in this regime.  Since the
classical tricritical point is at $\kappa_{{\rm cl}}=\frac{1}{2}$,
we shall investigate specifically the window
$0 \leq \kappa \leq 1$ of the frustration parameter.

For the case $s=\frac{1}{2}$, the model has been studied previously in
the same parameter range $0 \leq \kappa \leq 1$
\cite{Cabra:2011_honey,DJJF:2011_honeycomb,Bishop:2012_honey_circle-phase,Bishop:2015_honey_low-E-param}.
Each of these studies concurs on the finding that the classical
transition at $\kappa_{{\rm cl}}=\frac{1}{2}$ for the limiting case
$s \to \infty$ is split into two transitions for the case
$s=\frac{1}{2}$, one at $\kappa_{c_{1}} < \frac{1}{2}$ and the other
at $\kappa_{c_{2}} > \frac{1}{2}$.  Whereas LSWT provides the relatively crude
estimates $\kappa_{c_{1}} \approx 0.29$ and
$\kappa_{c_{2}} \approx 0.55$, the more powerful, and potentially more
accurate method of Schwinger-boson mean-field theory (SBMFT) gives the
estimates $\kappa_{c_{1}} \approx 0.41$ and
$\kappa_{c_{2}} \approx 0.6$ \cite{Cabra:2011_honey}.  These SBMFT
calculations also predict a quantum disordered phase in the
intermediate region $\kappa_{c_{1}} < \kappa < \kappa_{c_{2}}$, in
which a gap in the bosonic dispersion opens up.  These results are
broadly confirmed by high-order CCM calculations
\cite{DJJF:2011_honeycomb,Bishop:2012_honey_circle-phase,Bishop:2015_honey_low-E-param},
which yield the most accurate results to date for this case.  From
extensive calculations of a wide variety of low-energy parameters and
the triplet spin gap for the model, the best CCM estimates are
$\kappa_{c_{1}} = 0.45 \pm 0.02$ and $\kappa_{c_{2}} = 0.60 \pm 0.02$
\cite{Bishop:2015_honey_low-E-param}.  Furthermore, CCM calculations
of the plaquette valence-bond crystalline (PVBC) susceptibility
\cite{DJJF:2011_honeycomb} provide powerful evidence for the
intermediate paramagnetic phase to be a gapped state with PVBC order
over (almost all or) the entire region.  Furthermore, very recently,
the CCM has also been used to study a corresponding $AA$-stacked
honeycomb bilayer model
\cite{Bishop:2017_honeycomb_bilayer_J1J2J3J1perp} for the case
$s=\frac{1}{2}$, where each monolayer has the same bonds as here, but
now with the addition of an AFM NN interlayer coupling,
$J_{1}^{\perp}>0$.

In view of these interesting results for the $s=\frac{1}{2}$
$J_{1}$--$J_{2}$--$J_{3}$ model with $J_{3}=J_{2}$ and the known
results for the classical limit $(s \to \infty)$, it is clearly now
intriguing also to consider the cases $s=1$ and $s \geq 1$.  The only
results known to us are preliminary calculations of our own
\cite{Li:2016_honey_grtSpins}, again using the CCM.  In that earlier
work we calculated only the magnetic order parameter (out of the
complete set of low-energy parameters) for the two AFM quasiclassical
GS phases (i.e., the N\'{e}el and striped phases), for the cases
$s = 1, \frac{3}{2}, 2, \frac{5}{2}$, as well as their PVBC
susceptibilities.  Based only on the vanishing of the magnetic order
we found that the spin-1 case also had an intermediate non-classical
phase, with N\'{e}el ordering now disappearing at
$\kappa_{c_{1}}=0.485 \pm 0.005$ and striped order disappearing at
$\kappa_{c_{2}}=0.528 \pm 0.005$.  Just as for the $s=\frac{1}{2}$
case \cite{DJJF:2011_honeycomb} it was also found for the $s=1$ case
\cite{Li:2016_honey_grtSpins} that the transition at $\kappa_{c_{1}}$
appears to be of continuous type (and hence to be a candidate for
deconfined quantum criticality
\cite{Senthil:2004_Science_deconfinedQC,Senthil:2004_PRB_deconfinedQC},
since the Landau-Ginzburg-Wilson scenario cannot hold, as discussed in
more detail in Ref.\ \cite{DJJF:2011_honeycomb}), while that at
$\kappa_{c_{2}}$ appears to be of first-order type.  Unlike in the
$s=\frac{1}{2}$ case, however, where PVBC ordering seems to occur over
the entire range $\kappa_{c_{1}} < \kappa < \kappa_{c_{2}}$, for the
$s=1$ case PVBC ordering appears to be absent everywhere (or, at most,
to occur over only a very small part of the region).  For all cases
$s > 1$ studied in Ref.\ \cite{Li:2016_honey_grtSpins}, but based only
on calculations of the magnetic order parameter of the two AFM
quasiclassical phases, the quantum phase diagram appears to be similar
to the classical counterpart, i.e., with a direct first-order
transition from the N\'{e}el to the striped phase, but with a critical
value $\kappa_{c}(s)$ slightly greater than
$\kappa_{{\rm cl}}=\frac{1}{2}$.  Thus,
$\kappa_{c}(\frac{3}{2})=0.53 \pm 0.01$ and, for values
$s>\frac{3}{2}$, $\kappa_{c}(s)$ appears to approach the classical
value $\kappa_{c}(\infty)=0.5$ monotonically as $s$ is increased.

Our intention in the present paper is to add to these earlier
preliminary findings for the model with $s>\frac{1}{2}$ by calculating
for the cases $s=1$ and $s=\frac{3}{2}$ both a complete set of GS
low-energy parameters, to complement the calculations of the GS
magnetic order parameter, and the triplet spin gap.  In this sense we
parallel the development of the $s=\frac{1}{2}$ version of the model,
where the order parameter was first studied in Ref.\
\cite{DJJF:2011_honeycomb}, and only later were the more comprehensive
calculations of a complete set of low-energy parameters and the
triplet spin gap performed in Ref.\
\cite{Bishop:2015_honey_low-E-param}.  Specifically, we will present
results here in Sec.\ \ref{results_sec}, again using high-order CCM
calculations, for the spin stiffness coefficient $\rho_{s}$, the
zero-field (uniform) transverse magnetic susceptibility $\chi$, and
the spin gap $\Delta$.

\section{The coupled cluster method}
\label{ccm_section}
The CCM is nowadays regarded as providing one of the most accurate and
most flexible {\it ab initio} techniques of modern microscopic quantum
many-body theory (and see, e.g.,
Refs. \cite{Bishop:1987_ccm,Bishop:1991_TheorChimActa_QMBT,Bishop:1998_QMBT_coll,Bartlett:2007_ccm}).
The method is size-extensive and size-consistent at all levels of
approximate implementation.  Hence, it can be utilized from the outset
in the infinite system $(N \to \infty)$ limit, thereby obviating the
need for any finite-size scaling, such as is required by most
competing methods, and hence circumventing any associated source of
errors.  Furthermore, the method also exactly preserves both the
important Hellmann-Feynman theorem and the Goldstone linked cluster
theorem at every level of approximation.  These features ensure that
the CCM provides accurate and self-consistent sets of results for a
variety of both GS and excited-state (ES) parameters for the system
under study.  As is done here, the method can be implemented
computationally to high orders of approximation within well-defined
and well-understood truncation hierarchies, as outlined in more detail
below.  The results become exact as the order $n$ of the truncation
approaches infinity $(n \to \infty)$, and hence the {\it sole}
approximation ever made is to extrapolate the sequence of
approximants that we calculate for any specific physical parameter.

By now the CCM has been applied to a very large number of spin-lattice
systems, and we hence refer the reader to the extensive literature of
such applications (and see, e.g., Refs.\
\cite{Bishop:2015_honey_low-E-param,Li:2016_honey_grtSpins,Li:2016_honeyJ1-J2_s1,Zeng:1998_SqLatt_TrianLatt,Fa:2004_QM-coll,Bishop:2014_honey_XXZ_nmp14,Bishop:2017_SqLatt_XXZ,Bishop:2017_honeycomb_bilayer_J1J2J3J1perp}
and references cited therein) for full details.  We content ourselves
here with a very brief review of some of the most pertinent details
connected with the present applications.  A hallmark of the CCM is the
incorporation of the quantum correlations present in the exact GS or
ES wave functions via a very specific exponentiated form of
correlation operator acting on an appropriately chosen model (or
reference) state.  For GS quantities, such as the energy and magnetic
order parameter (i.e., the average local on-site magnetization) $M$,
we use (separately) here both the N\'{e}el and striped states shown in
Figs.\ \ref{model_pattern}(a) and \ref{model_pattern}(b),
respectively, as our choices of model state.  In order to calculate the lowest triplet ES energy gap
$\Delta$ for the two quasiclassical phases, a spin-1 reference state is created
from the above respective reference states by suitably exciting a
single spin.  For the calculations of the spin stiffness and magnetic
susceptibility the above GS reference states have to be suitably
modified by the imposition of a spin twist or a magnetic field,
respectively, as we now outline in more detail.

Firstly, the spin stiffness $\rho_{s}$ is a measure of the resistance
of the system to an imposed rotation of the order parameter by an
(infinitesimal) angle $\theta$ per unit length in a specific
direction.  If $E(\theta)$ is the GS energy as a function of the
imposed twist, then
\begin{equation}
\frac{E(\theta)}{N}=\frac{E(\theta=0)}{N}+\frac{1}{2}\rho_{s} \theta^{2} + O(\theta^{4})\,, \label{E_sStiff_theta_eq}
\end{equation}
where $\theta$ has the dimensions of inverse length.  Clearly,
magnetic LRO melts at the point where $\rho_{s}$ vanishes.  For the
N\'{e}el state, whose ordering wave vector
$\mathbf{Q}=\mathbf{\Gamma}=(0,0)$, clearly $\rho_{s}$ is independent
of the direction of the applied twist.  By contrast, for the striped
AFM state shown in Fig.\ \ref{model_pattern}(b), for which
$\mathbf{Q} = 2\pi/(\sqrt{3}d)\hat{x}$, the relevant direction to apply
the twist is the $x$ direction.  The corresponding twisted N\'{e}el
and twisted striped states \cite{Bishop:2015_honey_low-E-param} are then used
as CCM model states for the calculation of $\rho_{s}$ in both GS phases.
For the classical $(s \to \infty)$ version of the AFM
$J_{1}$--$J_{2}$--$J_{3}$ model under study on the honeycomb lattice,
with $J_{3}=J_{2}\equiv \kappa J_{1}$, the definition of Eq.\
(\ref{E_sStiff_theta_eq}) easily leads to the corresponding classical values of
$\rho_{s}$ in the two AFM phases,
\begin{equation}
\rho^{{\rm N\acute{e}el}}_{s;\,{\rm cl}}=\frac{3}{4}J_{1}(1-2\kappa)d^{2}s^{2}\,, \label{sStiff_neel_classical}
\end{equation}
and
\begin{equation}
\rho^{{\rm striped}}_{s;\,{\rm cl}}=\frac{3}{4}J_{1}(-1+2\kappa)d^{2}s^{2}\,. \label{sStiff_stripe_classical}
\end{equation}
both of which are positive in the respective regions of stability of
the two phases.  As expected, the spin stiffness of the system
vanishes precisely at the corresponding classical phase transition
point, $\kappa_{{\rm cl}}=\frac{1}{2}$.

Secondly, to calculate the transverse magnetic susceptibility when the
system is aligned in either AFM phase in the spin-space $z_{s}$
direction as in Fig.\ \ref{model_pattern}, we now place it in the
transverse magnetic field $\mathbf{h}=h\hat{x}_{s}$.  The Hamiltonian
$H=H(h=0)$ of Eq.\ (\ref{H_eq}) then becomes modified to
$H(h)=H(0)-h\sum_{k=1}^{N}s_{k}^{x}$, in units where the
gyromagnetic ratio $g\mu_{B}/\hbar=1$.  The spins, which were
previously aligned as in either Fig.\ \ref{model_pattern}(a) or Fig.\
\ref{model_pattern}(b), now cant at an angle $\phi=\phi(h)$ with
respect to their zero-field configurations along the $z_{s}$
direction.  These corresponding canted N\'{e}el and canted striped states
\cite{Bishop:2015_honey_low-E-param} are then used as CCM model states
for the calculation of the respective transverse magnetic
susceptibility, $\chi(h) = -N^{-1}{\rm d}^{2}E/{\rm d}h^{2}$, and its
zero-field limit, $\chi=\chi(0)$, in which we are interested.  In
practice we calculate $\chi$ from
\begin{equation}
\frac{E(h)}{N}=\frac{E(h=0)}{N}-\frac{1}{2}\chi h^{2} + O(h^{4})\,. \label{E_chi_theta_eq}
\end{equation}
for some suitably small value of $h$.  For the classical
$(s \to \infty)$ version of the AFM $J_{1}$--$J_{2}$--$J_{3}$ model
under study on the honeycomb lattice, with
$J_{3}=J_{2}\equiv \kappa J_{1}$, it is simple to calculate the
canting angle $\phi(h)$ for the two quasiclassical AFM phases.  The
definition of Eq.\ (\ref{E_chi_theta_eq}) thus easily yields the
corresponding classical values of $\chi$ for the two AFM phases,
\begin{equation}
\chi^{{\rm N\acute{e}el}}_{{\rm cl}}=\frac{1}{6J_{1}(1+\kappa)}\,, \label{chi_neel_classical}
\end{equation}
and
\begin{equation}
\chi^{{\rm striped}}_{{\rm cl}}=\frac{1}{2J_{1}(1+7\kappa)}\,,  \label{chi_stripe_classical}
\end{equation}
where both parameters are independent of $s$ at the classical level.
We note that the two values become equal $(=\frac{1}{9}J_{1}^{-1})$,
but nonzero, at the corresponding classical phase transition point,
$\kappa_{{\rm cl}}=\frac{1}{2}$.

We turn now to the choice of CCM approximation scheme.  Once a
suitable model state has been chosen, as outlined above, the
approximation simply involves the choice of which multispin-flip
configurations to retain in the CCM correlation operators.  A rather
general such approximation scheme is the so-called SUB$m$--$n$
hierarchy \cite{Fa:2004_QM-coll}.  At a given SUB$m$--$n$ level one
retains all multispin-flip configurations in the CCM correlation
operators that involve $m$ or fewer spin-flips spanning a range of no
more than $n$ contiguous sites on the lattice.  Each single spin-flip
is defined to require the action of a spin-raising operator
$s_{k}^{+} \equiv s_{k}^{x} + s_{k}^{y}$ acting once on the model ket
state (in local spin axes chosen so that on each site a passive
rotation has been performed to make each spin point downwards along
the negative $z_{s}$ axis).  Furthermore, a set of lattice sites
is defined to be contiguous if every site in the set is a NN (in a
specified geometry) to at least one other member of the set.  Clearly,
as both indices $m$ and $n$ become infinite, the approximation becomes
exact.
 
For spins with spin quantum number $s$, the maximum number of
spin-flips per site, defined as above, is $2s$.  Thus, when $m=2sn$,
the SUB$m$--$n$ scheme becomes equivalent to the so-called localized
lattice-animal-based subsystem (LSUB$n$) scheme in which all
multispin-flip configurations in the CCM correlation operator
expansions are retained that are defined over all distinct locales (or
lattice animals in the usual graph-theoretic sense) on the lattice
that comprise no more than $n$ contiguous sites.  Clearly, the
LSUB$n \equiv$ SUB$2sn$--$n$ scheme is only equivalent to the
SUB$n$--$n$ scheme for the case $s=\frac{1}{2}$.  In any such scheme
we utilize the (space- and point-group) symmetries of both the system
Hamiltonian and the CCM model state being used to reduce the set of
independent multispin-flip configurations retained at any given order
to a minimal number $N_{f}$.  At a given $n$th level of LSUB$n$
approximation, the number $N_{f}=N_{f}(n)$ is lowest for
$s=\frac{1}{2}$ and increases sharply as a function of $s$.  Because
$N_{f}(n)$ also increases rapidly (typically super-exponentially) with
the truncation index $n$, the most commonly used CCM truncation
hierarchy for spins $s > \frac{1}{2}$ is the SUB$n$--$n$ scheme, and
it is that scheme we employ here, just as in our previous work
\cite{Li:2016_honey_grtSpins} on this model for cases with $s \geq 1$.
In order to attain the higher values of the truncation index $n$
necessary for high accuracy, we also use massively parallel
supercomputing resources together with a specially tailored
computer-algebra package \cite{ccm_code} to derive and solve
\cite{Zeng:1998_SqLatt_TrianLatt} the corresponding sets of CCM bra-
and ket-state equations for both GS and ES quantities in the
SUB$n$--$n$ truncation scheme.

For the calculation of both the GS order parameter $M$ and the ES gap
parameter $\Delta$ we are restricted to SUB$n$--$n$ calculations with
$n \leq 12$ for the $s=\frac{1}{2}$ model and with $n \leq 10$ for the
corresponding models with $s \geq 1$.  For example, at the LSUB12
level for the calculation of $M$ in the case $s=\frac{1}{2}$, we have
$N_{f}(12)=103\,097$ $(250\,891)$ when the N\'{e}el (striped) state is
used as the CCM model state.  For comparison, at the SUB10--10 level
for calculations of $M$, we have corresponding numbers of fundamental
configurations $N_{f}(10)=219\,521$ $(552\,678)$ for the $s=1$ model
and $N_{f}(10)=461\,115$ $(1\,207\,202)$ for the corresponding
$s=\frac{3}{2}$ model.  Similarly at the LSUB12 level for the
calculation of $\Delta$ for the spin-$\frac{1}{2}$ model, the
corresponding numbers are $N_{f}(12)=182\,714$ $(465\,196)$ based on
the N\'{e}el (striped) state as CCM model state.  For comparison, at
the SUB10-10 level for calculations of $\Delta$, we have
$N_{f}(10)=244\,533$ $(642\,054)$ for the spin-1 model and
$N_{f}(10)=418\,164$ $(1\,123\,343)$ for the spin-$\frac{3}{2}$ model.

Due to the considerably reduced symmetries of both the twisted and
canted model states in both quasiclassical phases, the corresponding
SUB$n$--$n$ calculations of $\rho_{s}$ and $\chi$ can only be
performed at orders $n \leq 10$ for the case $s=\frac{1}{2}$ and
$n \leq 8$ for the cases $s \geq 1$.  Thus, for example, at the LSUB10
level for the calculation of $\rho_{s}$ in the case $s=\frac{1}{2}$,
we have $N_{f}(10)=347\,287$ when either the twisted N\'{e}el or the
twisted striped state is used as the CCM model state.  The corresponding
number for $\rho_{s}$ for the spin-1 model is $N_{f}(8) = 352\,515$ at
the SUB8--8 level.  Similarly, at the LSUB10 level for the calculation
of $\chi$ for the spin-$\frac{1}{2}$ model, we have
$N_{f}(10)=58\,537$ $(174\,692)$ when the canted N\'{e}el (canted
striped) state is used as the CCM model state.  Corresponding numbers
at the SUB8--8 level for $\chi$ for the spin-1 model are
$N_{f}(8)=59\,517$ $(177\,331)$.

As we have noted previously, once we have calculated CCM SUB$n$--$n$
approximants for any physical parameter the {\it only} approximation
that we ever make is the extrapolation to the (in principle) exact
limit, $n \to \infty$.  Although no exact such schemes are known
theoretically, by now a great deal of practical experience has been
built up from many applications to diverse models, so that we now have
a uniform set of simple extrapolation rules, one for each parameter,
that are applied consistently.  For example, for highly frustrated
spin-lattice models, particularly in cases where the system is close to
a QCP or where the magnetic order parameter $M$ is either zero or
close to zero, a well-tested scaling scheme for $M$ has been found
(and see, e.g., Refs.\
\cite{DJJF:2011_honeycomb,PHYLi:2012_honeycomb_J1neg,Bishop:2012_honeyJ1-J2,Bishop:2012_honey_circle-phase,Li:2012_honey_full,RFB:2013_hcomb_SDVBC,Li:2016_honey_grtSpins,Li:2016_honeyJ1-J2_s1,Bishop:2014_honey_XXZ_nmp14,Bishop:2017_SqLatt_XXZ}) to be
\begin{equation}
M(n) = \mu_{0}+\mu_{1}n^{-1/2}+\mu_{2}n^{-3/2}\,.   \label{M_extrapo_frustrated}
\end{equation}
Similar such schemes for the spin gap, spin stiffness and zero-field
transverse magnetic susceptibility all have a leading exponent of -1
(and see, e.g., Refs.\
\cite{Li:2016_honeyJ1-J2_s1,Bishop:2017_SqLatt_XXZ}),
\begin{equation}
\Delta(n) = d_{0}+d_{1}n^{-1}+d_{2}n^{-2}\,,   \label{Eq_spin_gap}
\end{equation}
\begin{equation}
\rho_{s}(n) = s_{0}+s_{1}n^{-1}+s_{2}n^{-2}\,,   \label{Eq_sstiff}
\end{equation}
and
\begin{equation}
\chi(n) = x_{0}+x_{1}n^{-1}+x_{2}n^{-2}\,.   \label{Eq_X}
\end{equation}
Clearly, in order to extract the corresponding extrapolants $\mu_{0}$
(for $M$), $d_{0}$ (for $\Delta$), $s_{0}$ (for $\rho_{s}$) and
$x_{0}$ (for $\chi$) from Eqs.\
(\ref{M_extrapo_frustrated})--(\ref{Eq_X}), respectively, we need to
input at least three different corresponding SUB$n$--$n$ approximants.
Further details on the choice of such SUB$n$--$n$ input sets are given
in Sec.\ \ref{results_sec}.

\section{Results}
\label{results_sec}
\begin{figure}
\mbox{
\subfigure[]{\includegraphics[height=7.5cm,angle=270]{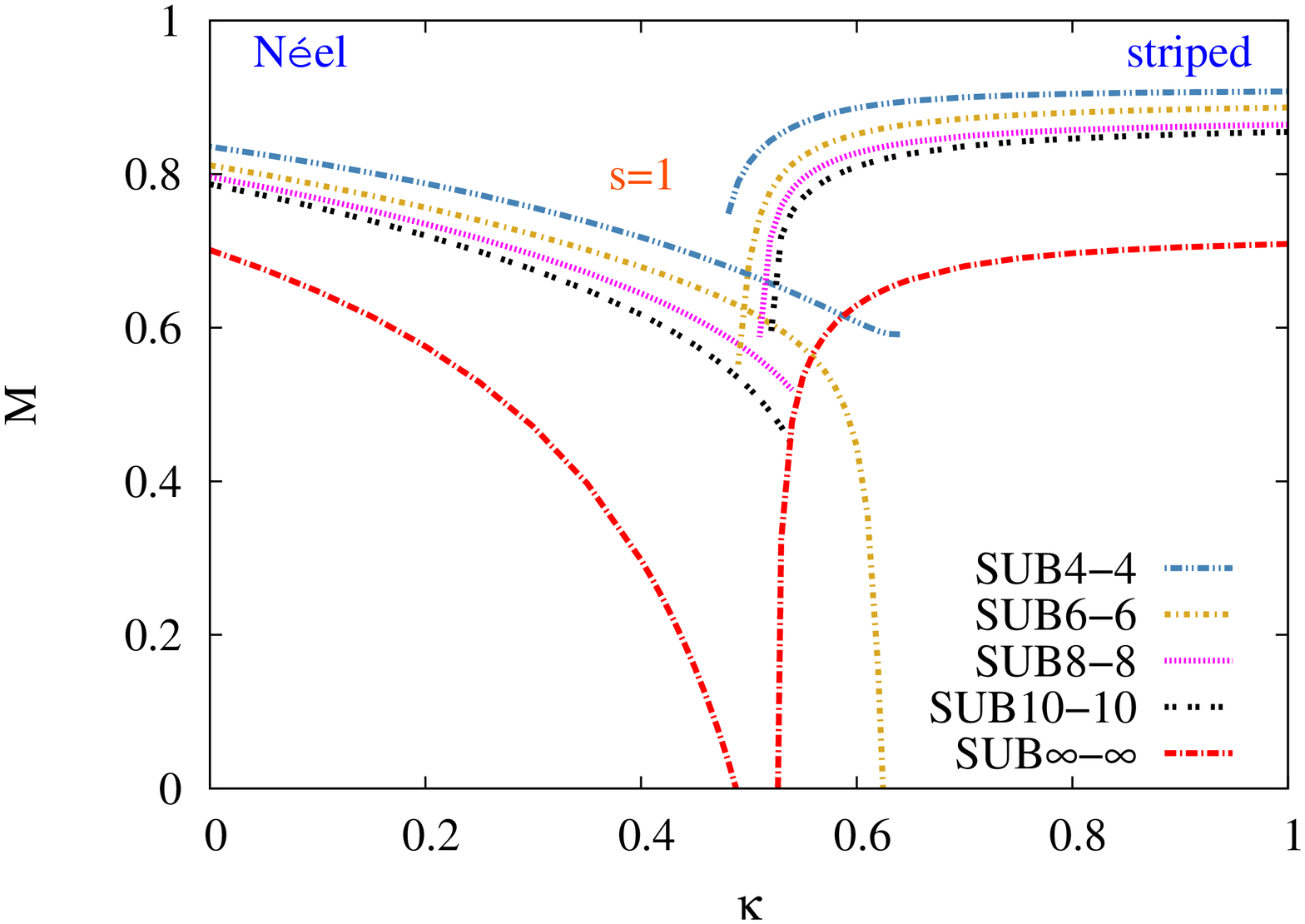}}
\quad \subfigure[]{\includegraphics[height=7.5cm,angle=270]{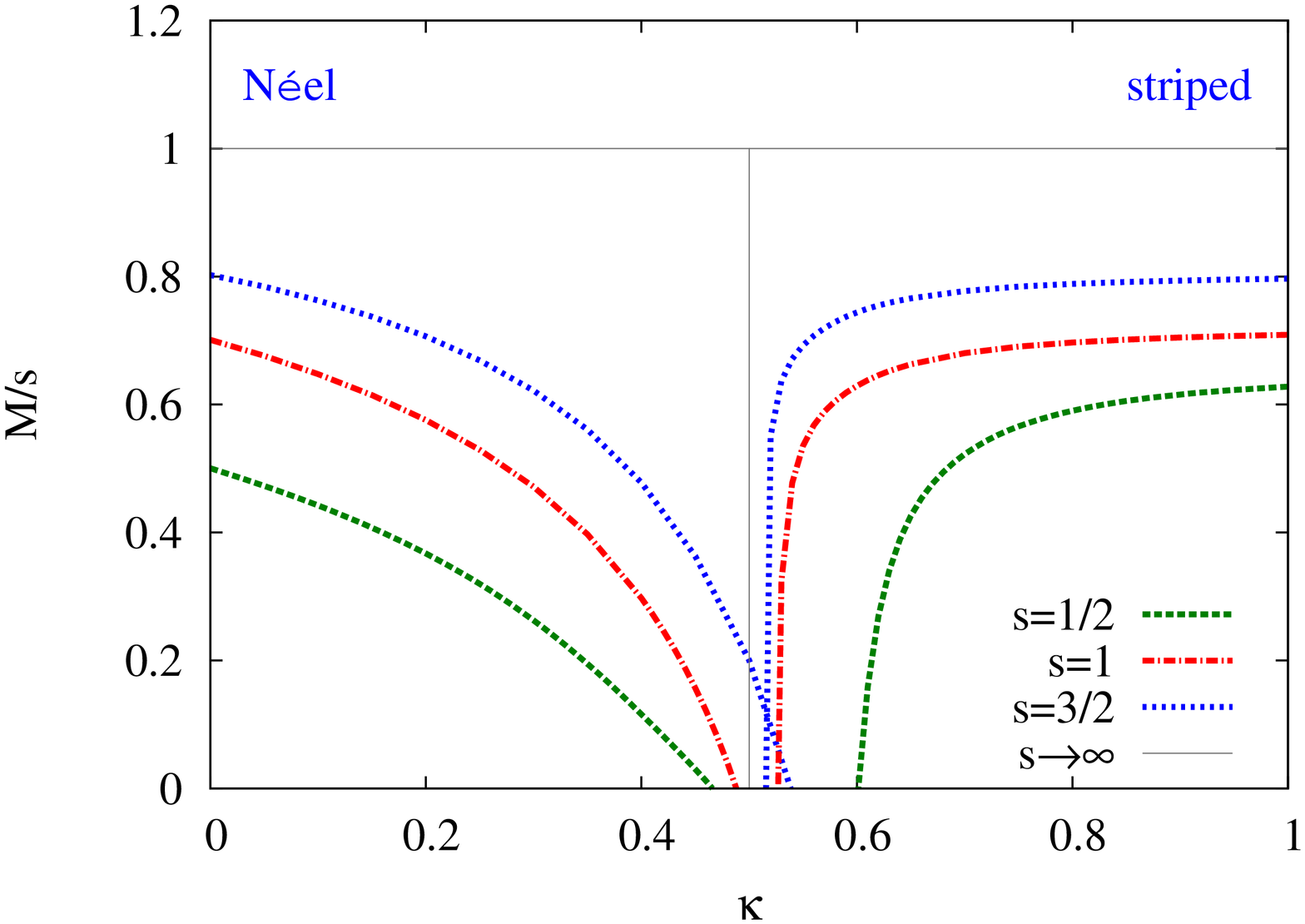}}
}
  \caption{CCM results for the GS magnetic order parameter $M$ for the $J_{1}$--$J_{2}$--$J_{3}$ model on the honeycomb lattice, with $J_{1}>0$ and $J_{3}=J_{2}\equiv\kappa J_{1}>0$, as a function of the frustration parameter $\kappa$, using both the N\'{e}el and striped states as the CCM model state.  (a) Results are shown for the $s=1$ case in SUB$n$--$n$ approximations with $n=4,6,8,10$, together with the respective SUB$\infty$--$\infty$ extrapolant based on Eq.\ (\ref{M_extrapo_frustrated}) and using this data set as input.  (b) Extrapolated (SUB$\infty$--$\infty$) results for $M/s$ are shown for the three cases $s=\frac{1}{2},1,\frac{3}{2}$.  In each case Eq.\ (\ref{M_extrapo_frustrated}) has been used, together with the corresponding SUB$n$--$n$ data sets $n=\{6,8,10,12\}$ for $s=\frac{1}{2}$ and $n=\{4,6,8,10\}$ for $s=1$ and $s=\frac{3}{2}$.  The classical $(s \to \infty)$ result is also shown.}
\label{M}
\end{figure}

We first show in Fig.\ \ref{M} our CCM results for the GS magnetic order parameter $M$.
In Fig.\ \ref{M}(a) we consider the spin-1 case, where we show both
the ``raw'' SUB$n$--$n$ data with $n=4,6,8,10$ (for the two cases
where the CCM model state is chosen to be either the N\'{e}el state or
the striped state) and the corresponding $n \to \infty$
(SUB$\infty$--$\infty$) extrapolations based on Eq.\
(\ref{M_extrapo_frustrated}) and using the SUB$n$--$n$ data set with
$n=\{4,6,8,10\}$ as input.  One sees that the extrapolated N\'{e}el
order parameter vanishes at a value $\kappa_{c_{1}} \approx 0.486$,
while the extrapolated striped order parameter similarly vanishes at a
value $\kappa_{c_{2}} \approx 0.527$.  As is usually the case, the
extrapolations are rather robust with respect to the choice of input
data.  For example, by comparing similar extrapolations using the
alternative data sets with $n=\{4,6,8\}$ and $n=\{6,8,10\}$, and by
making a more detailed error analysis of our results, we find that our
best estimates for the QCPs of the spin-1 model are
$\kappa_{c_{1}} = 0.485 \pm 0.005$ and
$\kappa_{c_{2}} = 0.528 \pm 0.005$, as already quoted in Sec.\
\ref{model_sec}.  These may be compared with the corresponding best
CCM estimates \cite{Bishop:2015_honey_low-E-param} of
$\kappa_{c_{1}} = 0.45 \pm 0.02$ and $\kappa_{c_{2}} = 0.60 \pm 0.02$
for the QCPs of the spin-$\frac{1}{2}$ model.

We note from Fig.\ \ref{M}(a) that each of the SUB$n$--$n$
approximants for $M$ terminates at some critical value of $\kappa$,
beyond which no real solution can be found for the corresponding CCM
equations.  Such critical values depend both on the model state used
and the order $n$ of the SUB$n$--$n$ approximation.  For the N\'{e}el
state there is an upper such critical value, while for the striped
state there is a corresponding lower critical value.  These CCM
termination points of the coupled sets of SUB$n$--$n$ equations are
typical of the method and are very well understood.  Thus, they are
simply a reflection of the corresponding QCP that delimits the region
of existence for the respective form of magnetic LRO pertaining
to that of the CCM model state being used.  For any specific finite
value $n$ of the SUB$n$--$n$ truncation index (and for a particular
phase under study), Fig.\ \ref{M}(a) clearly demonstrates that the CCM
solutions extend into the
unphysical regime beyond the actual ($n \to \infty$) QCP out to the respective
termination point.  As the truncation index $n$ is increased the
extent of the corresponding unphysical regime shrinks, and ultimately
disappears completely as $n \to \infty$ and the solution becomes exact.

The respective extrapolated CCM results for the scaled magnetic order
parameter, $M/s$, for both the N\'{e}el and striped states are
compared in Fig.\ \ref{M}(b) for the three cases $s=\frac{1}{2}$,
$s=1$ and $s=\frac{3}{2}$.  Results for $s=2$ and $s=\frac{5}{2}$ have
also been given previously in Ref.\ \cite{Li:2016_honey_grtSpins}.
One sees clearly that the evidence from our results for the magnetic
order parameter alone is that the intermediate phase exists only for
the two cases $s=\frac{1}{2}$ and $s=1$.  By contrast, for all cases
$s \geq \frac{3}{2}$ the order parameter results indicate a direct
transition between the N\'{e}el and striped phases at a QCP
$\kappa_{c}(s)$, just as occurs classically in the $s \to \infty$
limit.  For the case $s=\frac{3}{2}$ shown in Fig.\ \ref{M}(b), we
find $\kappa_{c}(\frac{3}{2}) = 0.517 \pm 0.004$, based on the order
parameter results alone.  It has also been observed (and see Ref.\
\cite{Li:2016_honey_grtSpins} for further details) that
$\kappa_{c}(s)$ appears to approach monotonically the classical value
$\kappa_{{\rm cl}} \equiv \kappa_{c}(\infty) = 0.5$ as $s$ is
increased further.

\begin{figure}
\mbox{
\subfigure[]{\includegraphics[height=3.7cm]{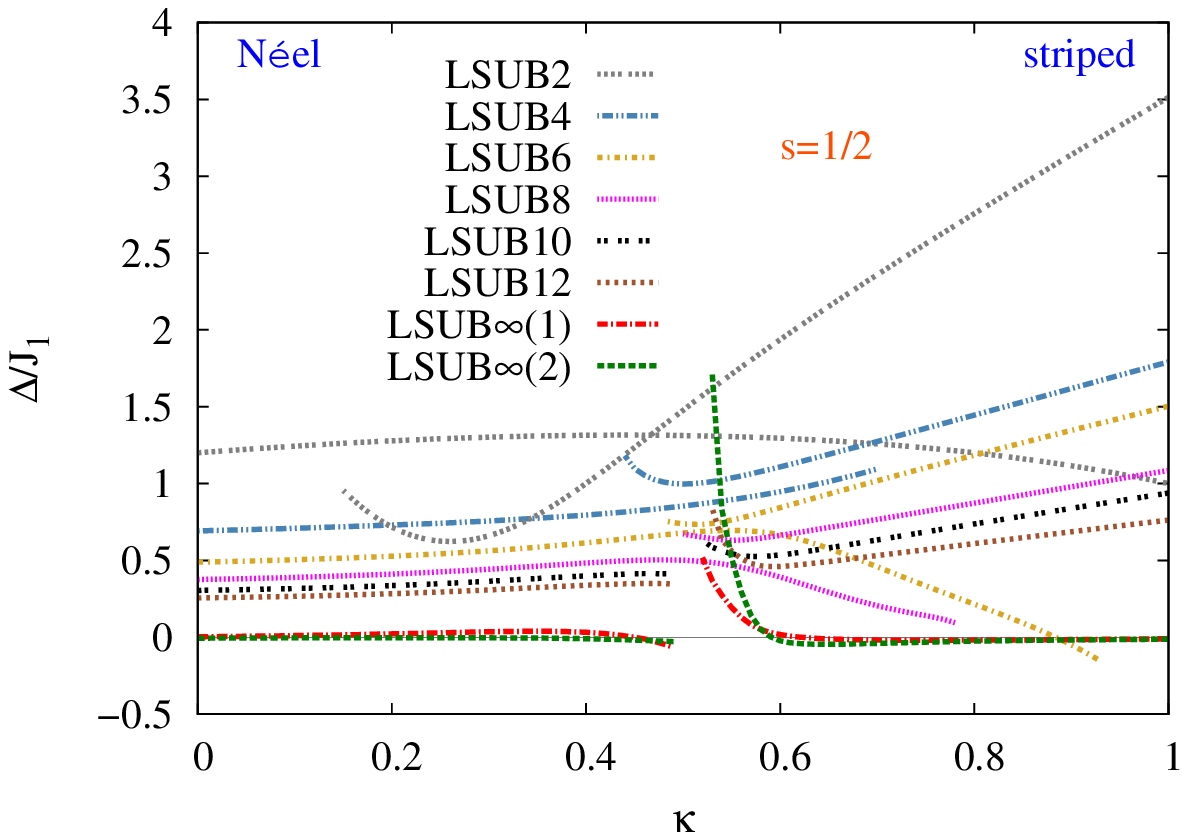}}
\subfigure[]{\includegraphics[height=3.7cm]{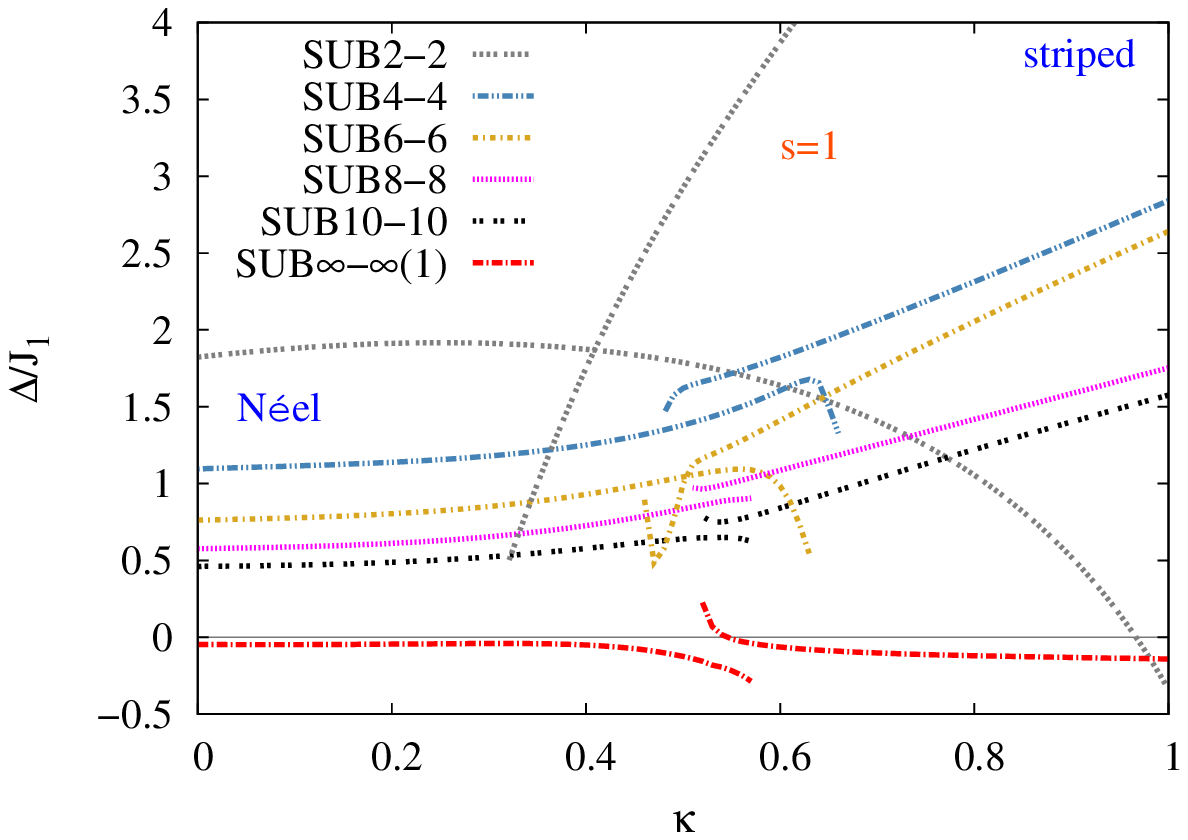}}
\subfigure[]{\includegraphics[height=3.7cm]{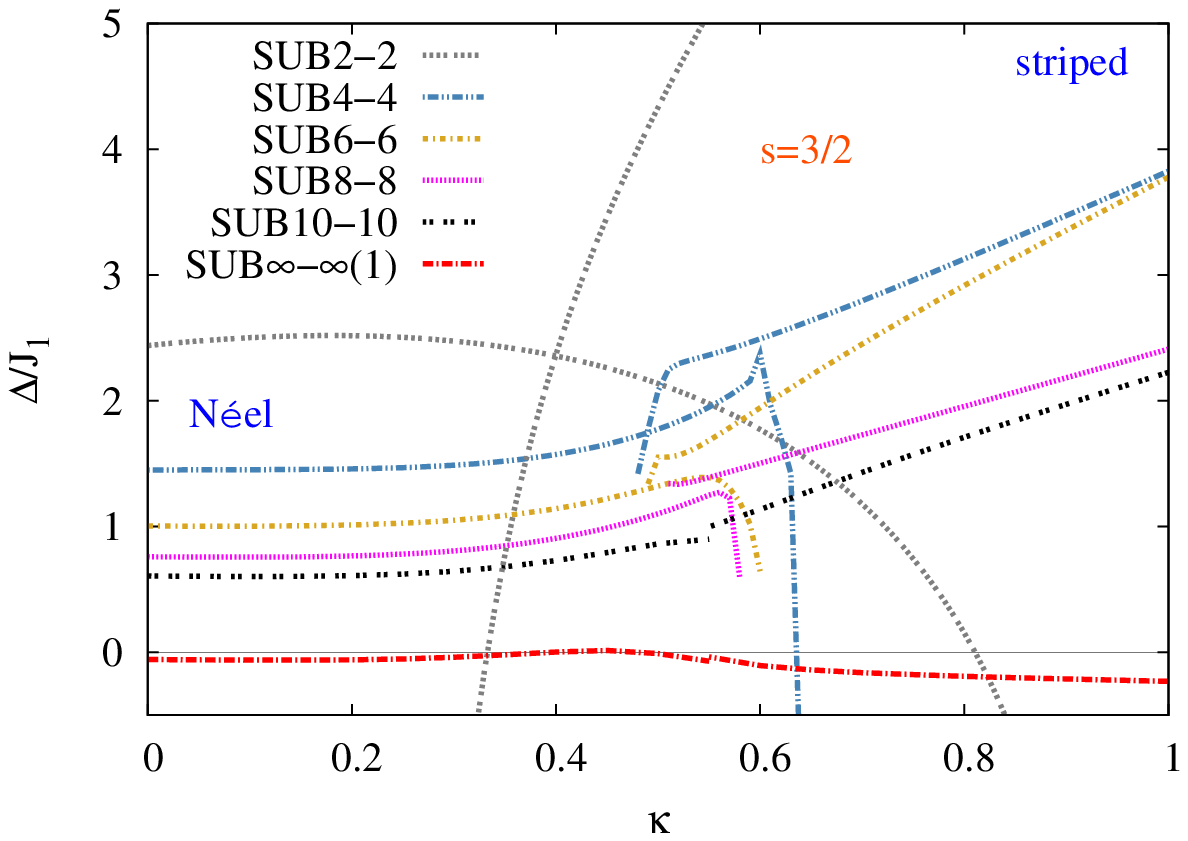}}
}
  \caption{CCM results for the scaled triplet spin gap $\Delta/J_{1}$ for the $J_{1}$--$J_{2}$--$J_{3}$ model on the honeycomb lattice, with $J_{1}>0$ and $J_{3}=J_{2}\equiv\kappa J_{1}>0$, as a function of the frustration parameter $\kappa$, using both the N\'{e}el and striped states as the CCM model state, for the three cases (a) $s=\frac{1}{2}$, (b) $s=1$, and (c) $s=\frac{3}{2}$.  Results are shown in SUB$n$--$n$ ($\equiv$ LSUB$n$ for $s=\frac{1}{2}$ only) approximations with $n=2,4,6,8,10$ in each case and also with $n=12$ for the case $s=\frac{1}{2}$ only.  Extrapolated SUB$\infty$--$\infty(i)$ results are also shown, based in each case on Eq.\ (\ref{Eq_spin_gap}) and the respective data sets $n=\{2,6,10\}$ for $i=1$ and $n=\{4,8,12\}$ for $i=2$ (in the case $s=\frac{1}{2}$ only).}
\label{Egap_fig}
\end{figure}

In order to provide further information about the QCPs observed from
the magnetic order parameter results in Fig.\ \ref{M}, we now first
turn our attention to the triplet spin gap $\Delta$.  We display our
corresponding CCM results for $\Delta/J_{1}$ as a function of $\kappa$
in Figs.\ \ref{Egap_fig}(a), \ref{Egap_fig}(b) and \ref{Egap_fig}(c)
respectively for the three cases $s=\frac{1}{2}$, $s=1$ and
$s=\frac{3}{2}$.
In each case raw SUB$n$--$n$ and extrapolated (SUB$\infty$--$\infty$)
results are shown based on both the N\'{e}el and striped collinear AFM
states used separately as our CCM model state.  Once again we observe
termination points for the excited-state CCM equations, in complete
analogy to those discussed above for the corresponding GS equations in
connection with the magnetic order parameter results shown in Fig.\
\ref{M}.

At this point we should note that a $(4m-2)/4m$ staggering effect,
where $m \in \mathbb{Z}^{+}$ is a positive integer, has been observed
previously
\cite{Bishop:2012_honeyJ1-J2,RFB:2013_hcomb_SDVBC,Li:2016_honeyJ1-J2_s1,Bishop:2017_honeycomb_bilayer_J1J2J3J1perp}
in CCM SUB$n$--$n$ sequences of results (with $n$ an even integer) for
a variety of physical parameters on frustrated honeycomb lattices.
Such staggering occurs when the SUB$n$--$n$ subsequence of results
for some parameter with $n=(4m-2)$ is differentially offset (or staggered) with
respect to the corresponding sub-sequence with $n=4m$.  Both
sub-sequences still obey an extrapolation scheme of the same sort
(i.e., with the same leading exponent), but the coefficients are not
identical for both cases.  Such an effect is well known both in
perturbation theory and in the CCM for all lattices as a corresponding
$(2m-1)/2m$ (i.e., odd/even) staggering.  It is precisely for this
reason that we do not show SUB$n$--$n$ results here also for odd
values of $n$, since one should not mix odd and even sub-sequences in
a single extrapolation.  What is novel for honeycomb-lattice models is
an {\it additional} staggering in the even-order series of terms for
some physical observables between those with $n=(4m-2)$ and those with
$n=4m$.  It has been postulated \cite{Li:2016_honeyJ1-J2_s1} that such
an additional staggering can be attributed to the fact that the
honeycomb lattice is non-Bravais, comprising two interlocking
triangular Bravais lattices, on each of which the usual $(2m-1)/2m$
staggering is observed.

While such a $(4m-2)/4m$ staggering is not appreciable in the results
shown in Fig.\ \ref{M} for $M$, it is visibly apparent in the results
shown in Fig.\ \ref{Egap_fig} for $\Delta$, particularly in the
striped phase, but also, albeit to a lesser extent, in the N\'{e}el
phase.  For this reason in our extrapolations for $\Delta$ based on
Eq.\ (\ref{Eq_spin_gap}) we take care not to mix SUB$n$--$n$ terms in
the input set with $n=(4m-2)$ and those with $n=4m$.  Thus, for the
case $s=\frac{1}{2}$ alone, shown in Fig.\ \ref{Egap_fig}(a), where we
have SUB$n$--$n$ ($\equiv {\rm LSUB}n$ in this case) results with
$n \leq 12$, we compare the extrapolation based on $n=\{2,6,10\}$ with
that based on $n=\{4,8,12\}$.  The agreement between the two is
excellent in both stable quasiclassical phases, and again demonstrates
the robustness of our extrapolation procedures.  The only appreciable
difference occurs precisely in the region of the intermediate
paramagnetic state near to the boundary with the striped state.  Since
this is precisely the unphysical part of the region accessible with
the striped CCM model state, it is where the associated errors are
expected to be the largest and also the most difficult to estimate.
What is clear, however, is that the evidence points strongly towards a
gapped ($\Delta > 0$) state for the $s=\frac{1}{2}$ case in this
intermediate region.  For the corresponding cases with $s=1$ and
$s=\frac{3}{2}$ shown in Figs.\ \ref{Egap_fig}(b) and
\ref{Egap_fig}(c) respectively, we are only able to perform
SUB$n$--$n$ calculations for $\Delta$ with $n \leq 10$.  Hence, for
these two cases, we only show the extrapolations for $\Delta$ based on
Eq.\ (\ref{Eq_spin_gap}) with the data set $n=\{2,6,10\}$ used as
input.

It is very gratifying to note firstly from Fig.\ \ref{Egap_fig} that
for both quasiclassical magnetic phases our extrapolated values for
$\Delta$ are compatible with being zero, within very small numerical
errors, over the ranges $\kappa < \kappa_{c_{1}}$ and
$\kappa > \kappa_{c_{2}}$ (with QCPs at $\kappa_{c_{1}}$ and
$\kappa_{c_{2}}$ as determined from Fig.\ \ref{M} by the points where
$M \to 0$) for the two cases $s=\frac{1}{2}$ and $s=1$, and over the
entire range $0 \leq \kappa \leq 1$ for the case $s=\frac{3}{2}$.
This is exactly as expected for the case of magnetic LRO, where the
low-energy magnon excitations are gapless.  The very small negative values shown for parts of the quasiclassical regions in Fig.\ \ref{Egap_fig} are clearly just an artefact of the numerical extrapolations.  Indeed, their magnitude gives us a good independent check on the overall level of accuracy of our results.  

For the case $s=\frac{1}{2}$, shown in Fig.\ \ref{Egap_fig}(a), there is
clear evidence that in the intermediate paramagnetic regime the GS
phase is gapped (i.e., with $\Delta > 0$), at least in the regime near
$\kappa_{c_{2}}$, where the excited-state LSUB$n$ calculations based
on the striped state as CCM model state extend considerably into the
``unphysical regime'' before the corresponding termination point
(discussed previously for the GS solutions) is reached.  
By contrast,
the excited-state LSUB$n$ calculations based on the N\'{e}el state
terminate for the higher values of $n$ shown, and hence also for the
LSUB$\infty(i)$ extrapolants, only just beyond the value
$\kappa_{c_{1}}$ at which the N\'{e}el magnetic order parameter
vanishes, as in Fig.\ \ref{M}.  Correspondingly, it is more difficult,
on the basis of the evidence from $\Delta$ alone, to state
categorically that the intermediate state is gapped over the entire
range $\kappa_{c_{1}} < \kappa < \kappa_{c_{2}}$ for the
spin-$\frac{1}{2}$ model.  Nevertheless, there is clear evidence that
at the QCP $\kappa_{c_{2}}$ where the striped order melts, the emergent
paramagnetic state {\it is} gapped, and that this gapped state
persists over most (if not all) of the regime
$\kappa_{c_{1}} < \kappa < \kappa_{c_{2}}$.

Turning next to the case $s=\frac{3}{2}$, shown in Fig.\
\ref{Egap_fig}(c), it is clear that the extrapolant for $\Delta$ is zero
(within numerical errors) over the entire range of values of the
frustration parameter $\kappa$ shown.  For the highest-order
SUB$n$--$n$ calculations shown with $n=10$, the two results, based
separately on the N\'{e}el and striped states as CCM model states,
terminate at essentially the same point.  Furthermore, the two
corresponding SUB10-10 results for $\Delta$, shown in Fig.\
\ref{Egap_fig}(c), lie on a single curve that is continuous and smooth,
within very small uncertainties.  The same is also true for the
SUB$\infty$--$\infty(1)$ extrapolant based on Eq.\ (\ref{Eq_spin_gap})
and the input data set $n=\{2,6,10\}$.  These results entirely
corroborate our findings from Fig.\ \ref{M} for the order parameter,
that there is a direct transition in the spin-$\frac{3}{2}$ case
between the two quasiclassical collinear AFM states.

By contrast with the very clear results for $\Delta$ for both the
$s=\frac{1}{2}$ case shown in Fig.\ \ref{Egap_fig}(a) and the
$s=\frac{3}{2}$ case shown in Fig.\ \ref{Egap_fig}(c), the results in
Fig.\ \ref{Egap_fig}(b) for the $s=1$ case are more open to interpretation.
For example, in this case, even the two highest-order excited-state
SUB$n$--$n$ solutions shown (i.e., with $n=10$), based separately on
both the N\'{e}el and striped states as CCM model state, now show some
(possibly unphysical) overlap region where both solutions exist.
However, unlike the $s=\frac{3}{2}$ case, the two SUB10-10 solutions
in this regime, while close to one another, do not smoothly overlap,
as is also the case for the two corresponding SUB$\infty$--$\infty(1)$
extrapolants.  It seems unlikely that these differences can be
attributed to numerical errors, although we cannot wholly rule this
out.  On balance, the evidence from the results in Fig.\
\ref{Egap_fig}(b) for $\Delta$, corroborate our finding from Fig.\
\ref{M}(b) that there is a small regime for the spin-1 model between
the N\'{e}el and striped phases where the stable GS phase is
paramagnetic.  Furthermore, it also seems reasonably clear that near
the QCP where this intermediate state meets the striped state, the
former is also gapped, as in the spin-$\frac{1}{2}$ case.  From Fig.\
\ref{Egap_fig}(b) the point where the extrapolant shown for $\Delta$
becomes positive, viz., at $\kappa \approx 0.546$ is also very close
to the value $\kappa_{c_{2}} \approx 0.528$ from Fig.\ \ref{M}(b) at
which the striped magnetic LRO melts.

Finally, we turn our attention to the two remaining GS low-energy
parameters (viz., the spin stiffness coefficient $\rho_{s}$ and the
zero-field transverse magnetic susceptibility $\chi$).  Since the
results for the cases $s \geq \frac{3}{2}$ are already conclusive
(i.e., that there is a direct first-order transition in each case
between the N\'{e}el and striped AFM phases), we concentrate
henceforward on comparing the two cases $s=\frac{1}{2}$ and $s=1$.
Due to the considerably reduced symmetries of both the twisted and
canted (N\'{e}el and striped) CCM model states required to
calculate $\rho_{s}$ and $\chi$, respectively, we are now only able to
perform LSUB$n$ calculations for $s=\frac{1}{2}$ with $n \leq 10$ and
SUB$n$--$n$ calculations for $s \geq 1$ with $n \leq 8$, for both
parameters.  This can be compared with the corresponding cases
$n \leq 12$ and $n \leq 10$ respectively, for the calculations of $M$
and $\Delta$.

\begin{figure}
\mbox{
\subfigure[]{\includegraphics[height=7.5cm,angle=270]{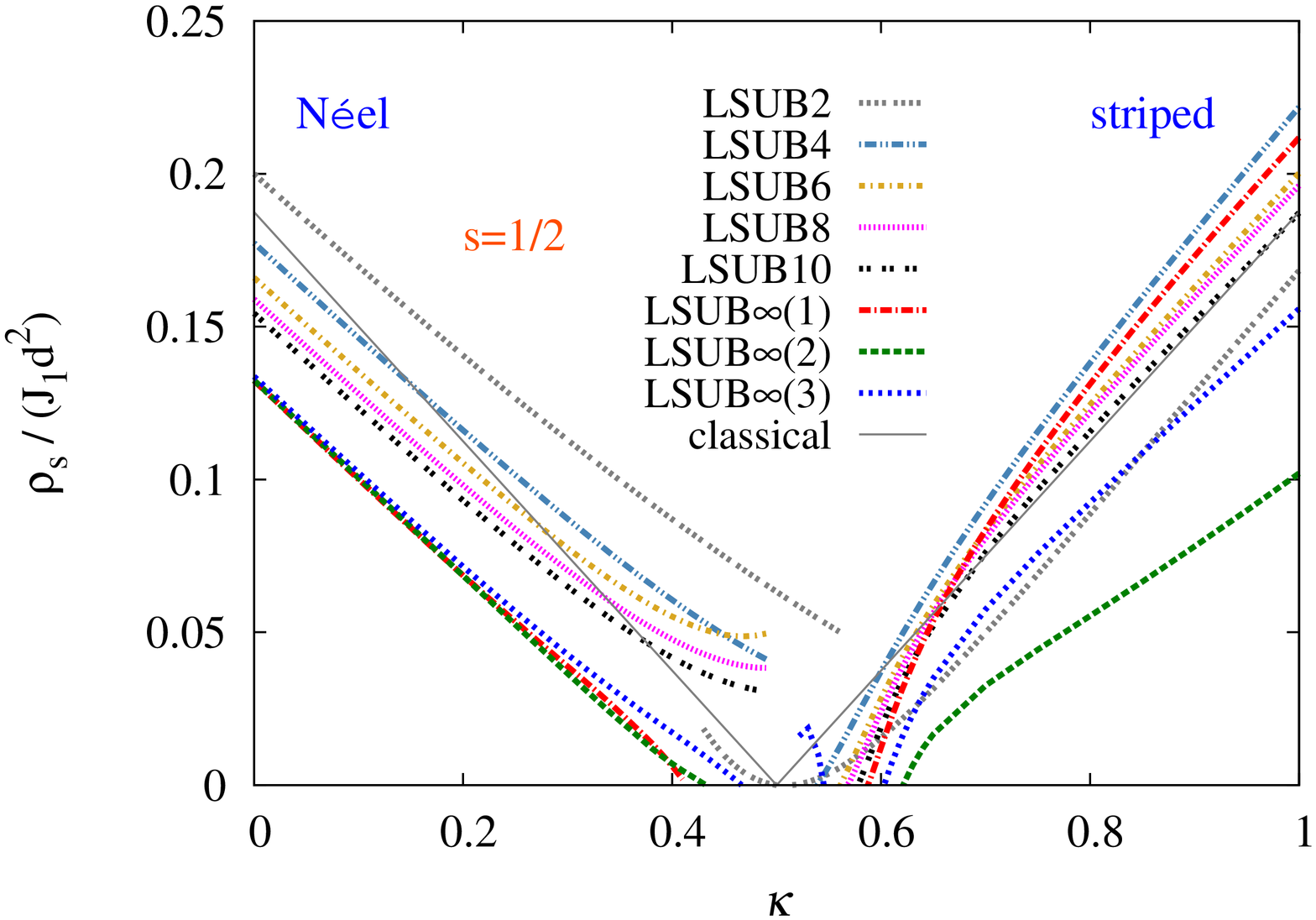}}
\quad \subfigure[]{\includegraphics[height=7.5cm,angle=270]{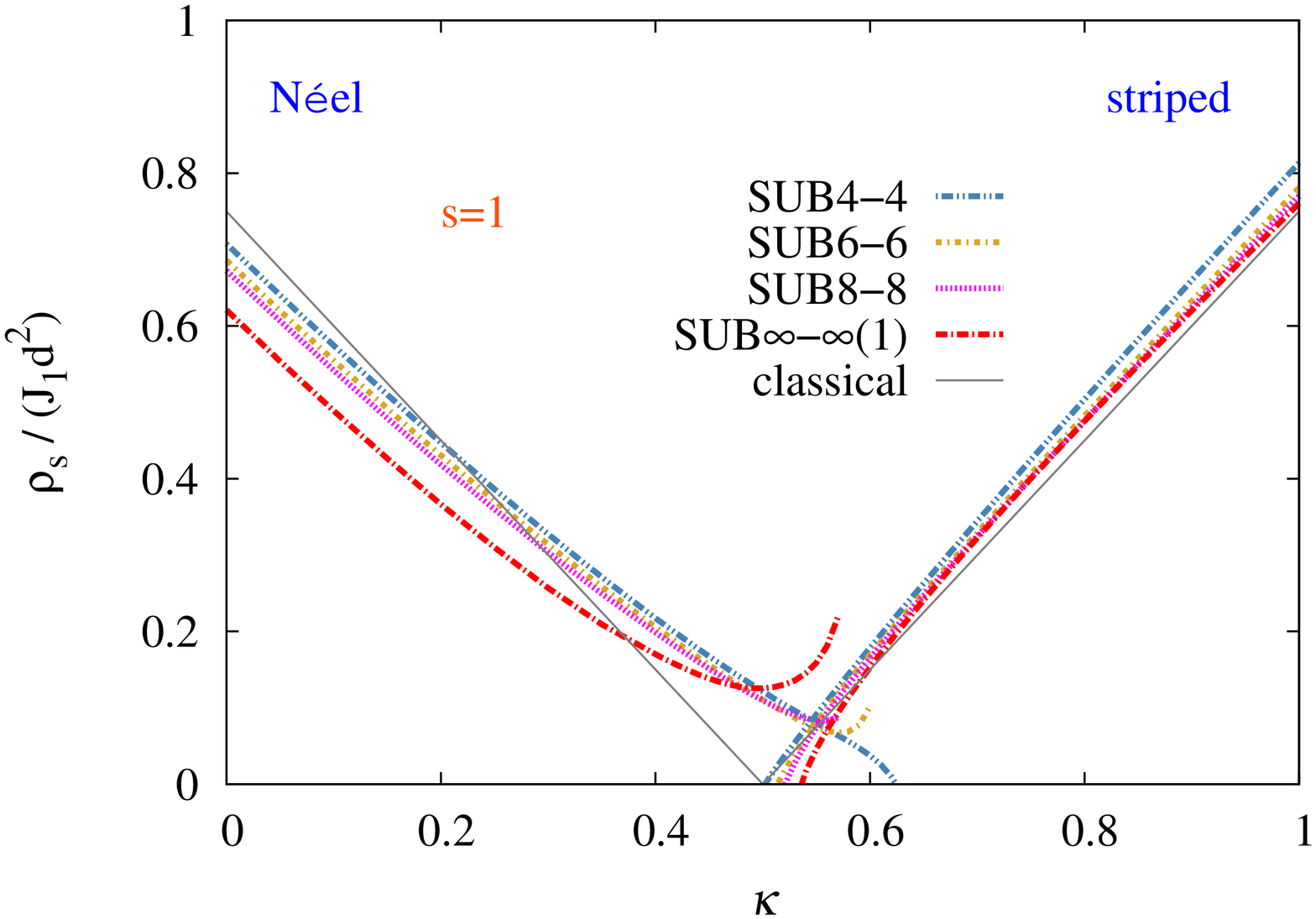}}
}
\caption{CCM results for the scaled spin stiffness
  $\rho_{s}/(J_{1}d^{2})$ for the $J_{1}$--$J_{2}$--$J_{3}$ model on
  the honeycomb lattice, with $J_{1}>0$ and
  $J_{3}=J_{2}\equiv\kappa J_{1}>0$, as a function of the frustration
  parameter $\kappa$, using both the twisted N\'{e}el and twisted striped
  states as the CCM model state, for the two cases (a) $s=\frac{1}{2}$
  and (b) $s=1$.  Results are shown in SUB$n$--$n$ ($\equiv$ LSUB$n$
  for $s=\frac{1}{2}$ only) approximations with $n=2,4,6,8,10$ for
  $s=\frac{1}{2}$ and $n=4,6,8$ for $s=1$.  Extrapolated
  SUB$\infty$--$\infty(i)$ results are also shown, based in each case
  on Eq.\ (\ref{Eq_sstiff}) and the respective data sets $n=\{4,6,8\}$
  for $i=1$, $n=\{6,8,10\}$ for $i=2$ and $n=\{2,6,10\}$ for $i=3$ (the latter two in the case $s=\frac{1}{2}$
  only).  The classical $(s \to \infty)$ results from Eqs.\ (\ref{sStiff_neel_classical}) and (\ref{sStiff_stripe_classical}) are also shown, using the values $s=\frac{1}{2}$ and $s=1$ in panels (a) and (b), respectively.}
\label{sStiff}
\end{figure}

In Fig.\ \ref{sStiff} we display our CCM results for $\rho_{s}$ (in units
of $J_{1}d^{2}$), using both the twisted N\'{e}el and twisted canted
states separately as model states.
The ``raw'' LSUB$n$ results for the spin-$\frac{1}{2}$ case, shown in
Fig.\ \ref{sStiff}(a) are seen to be clearly different in character
for the N\'{e}el state from their counterparts for the striped state.
Thus, firstly, in the former case, the curves flatten and seem to
acquire zero slope before or near their termination points, while the
latter become steeper near their termination points.  This is rather
strong evidence for the QCP at $\kappa_{c_{1}}$ where N\'{e}el order
melts being of continuous type, while that at $\kappa_{c_{2}}$ where
striped order melts is of first-order type.  Secondly, whereas the
LSUB$n$ results for the N\'{e}el state show no perceivable $(4m-2)/4m$
staggering of the sort discussed above, such an effect is clearly seen
in their counterparts for the striped state.

This latter difference is also clearly reflected in the behaviour of
the extrapolated values.  Thus, in Fig.\ \ref{sStiff}(a) we display
three different extrapolants LSUB$\infty(i)$, each of which is based
on the scheme of Eq.\ (\ref{Eq_sstiff}), but using the three different
input data sets $n=\{4,6,8\}$ for $i=1$, $n=\{6,8,10\}$ for $i=2$ and
$n=\{2,6,10\}$ for $i=3$.  Clearly, for the N\'{e}el phase, the two
extrapolants with $i=1$ and $i=2$ are nearly identical, thereby
demonstrating both the lack of any appreciable $(4m-2)/4m$ staggering
and the robustness of our extrapolation procedure using Eq.\
(\ref{Eq_sstiff}).  Interestingly, the only noticeable difference
occurs in a very small region near $\kappa_{c_{1}}$ where the
LSUB$\infty(2)$ curve, which utilizes higher-order LSUB$n$
approximants than the LSUB$\infty(1)$ curve (and which hence a {\it
  priori} is expected to be more accurate), now more closely reflects
the continuous nature of the transition with $\rho_{s}$ approaching
zero with zero slope.  Even so, the values for $\kappa_{c_{1}}$
obtained from the two extrapolations, viz.,
$\kappa_{c_{1}} \approx 0.411$ from LSUB$\infty(1)$ and
$\kappa_{c_{1}} \approx 0.433$ from LSUB$\infty(2)$, are in close
agreement.  By contrast, the two LSUB$\infty(i)$ extrapolants for the
striped phase are not at all in agreement with one another for essentially
all values of $\kappa$ shown.  This is clearly a reflection of the now
marked $(4m-2)/4m$ staggering, which is clearly perceived by visual
inspection of the corresponding LSUB$n$ curves, shown in Fig.\
\ref{sStiff}(a), for the striped phase.  The corresponding values now
obtained for $\kappa_{c_{2}}$ (i.e., the points where $\rho_{s}$ for
the striped phase vanishes) are $\kappa_{c_{2}} \approx 0.588$ from
LSUB$\infty(1)$ and $\kappa_{c_{2}} \approx 0.621$ from
LSUB$\infty(2)$.

Obviously, once such $(4m-2)/4m$ staggering effects have been
detected, as above, one should base the extrapolations only on LSUB$n$
subsequences with $n=(4m-2)$ alone or with $n=4m$ alone.  For the
spin-$\frac{1}{2}$ case, where we have LSUB$n$ results for $\rho_{s}$
with $n \leq 10$, we can hence use an extrapolation based on Eq.\
(\ref{Eq_sstiff}) and input data set $n=\{2,6,10\}$, which should
circumvent any complications of staggering.  These are just the
LSUB$\infty(3)$ results shown in Fig.\ \ref{sStiff}(a).  Again, on the
N\'{e}el side, all three extrapolations are in good agreement with
each other except close to the QCP at $\kappa_{c_{1}}$, whereas on the
striped side the LSUB$\infty(3)$ curve should now be clearly taken as
our preferred result.  The corresponding locations of the two QCPs
from the LSUB$\infty(3)$ extrapolation are
$\kappa_{c_{1}} \approx 0.466$ and $\kappa_{c_{2}} \approx 0.603$.
Both are in complete agreement with our previously quoted best CCM
estimates \cite{Bishop:2015_honey_low-E-param} of
$\kappa_{c_{1}} \approx 0.45 \pm 0.02$ and
$\kappa_{c_{2}} \approx 0.60 \pm 0.02$.

Turning now to the corresponding spin-1 case, shown in Fig.\
\ref{sStiff}(b), we only have SUB$n$--$n$ results with $n \leq 8$, and
hence we only display the SUB$\infty$--$\infty(1)$ extrapolant based
on Eq.\ (\ref{Eq_sstiff}) and the $n=\{4,6,8\}$ input data set.  Clearly,
on the N\'{e}el side the extrapolated value for $\rho_{s}$ does not
vanish, but instead displays a minimum at a value
$\kappa \approx 0.495$.  This is clearly a shortcoming of the
extrapolation, which would presumably disappear if we had higher-order
SUB$n$--$n$ approximants available (e.g., with $n=10$).  Nevertheless,
even this limited extrapolant is showing clear evidence for the
corresponding QCP being continuous, just as in the spin-$\frac{1}{2}$
case.  By contrast, on the striped side, where the QPT is of
first-order type, the extrapolated result for $\rho_{s}$ vanishes at a
value $\kappa \approx 0.562$, in reasonable agreement with the
previously quoted best CCM estimate \cite{Li:2016_honey_grtSpins} of
$\kappa_{c_{2}} \approx 0.528 \pm 0.005$, obtained from the vanishing
of the N\'{e}el order parameter.  In summary, our present results for
$\rho_{s}$ for the spin-1 model reinforce our previous conclusions,
while not adding significantly to their accuracy due to the fact that
calculations for $\rho_{s}$ can only be performed to lower orders (for
comparable computing resources) than for $M$.

\begin{figure}
\mbox{
\subfigure[]{\includegraphics[height=7.5cm,angle=270]{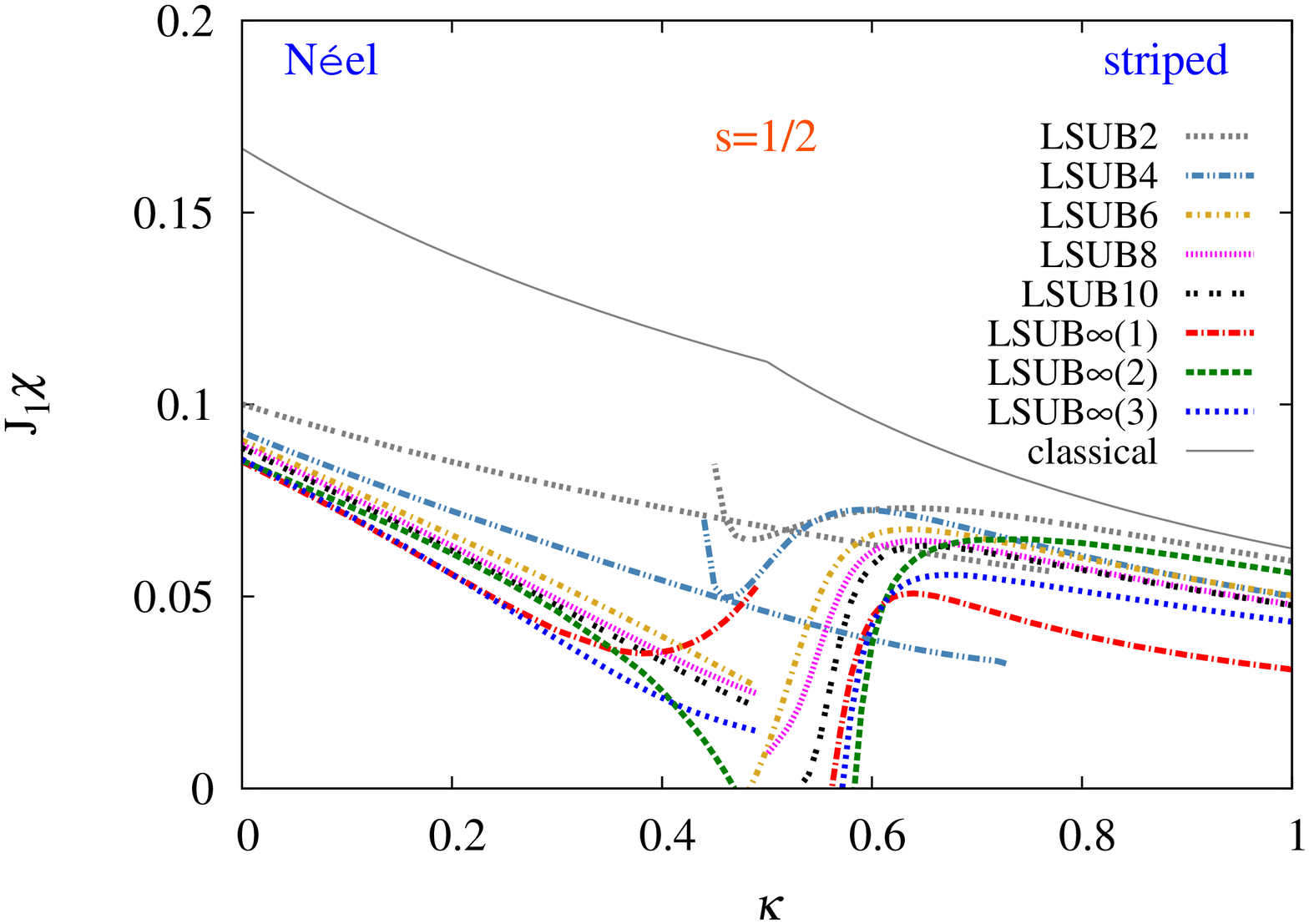}}
\quad \subfigure[]{\includegraphics[height=7.5cm,angle=270]{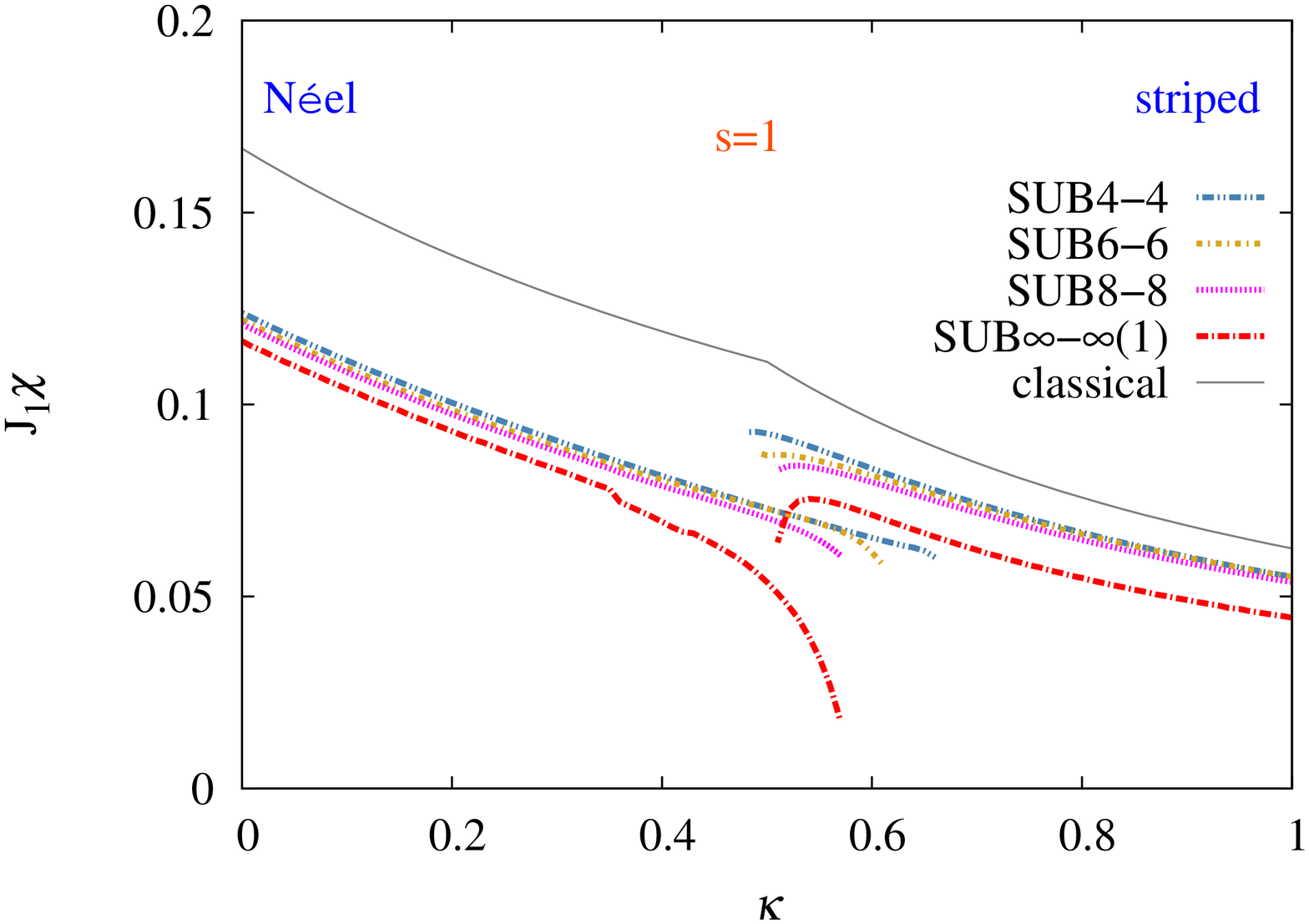}}
}
  \caption{CCM results for the scaled zero-field, transverse magnetic susceptibility $J_{1}\chi$ for the $J_{1}$--$J_{2}$--$J_{3}$ model on
  the honeycomb lattice, with $J_{1}>0$ and
  $J_{3}=J_{2}\equiv\kappa J_{1}>0$, as a function of the frustration
  parameter $\kappa$, using both the canted N\'{e}el and canted striped
  states as the CCM model state, for the two cases (a) $s=\frac{1}{2}$
  and (b) $s=1$.  Results are shown in SUB$n$--$n$ ($\equiv$ LSUB$n$
  for $s=\frac{1}{2}$ only) approximations with $n=2,4,6,8,10$ for
  $s=\frac{1}{2}$ and $n=4,6,8$ for $s=1$.  Extrapolated
  SUB$\infty$--$\infty(i)$ results are also shown, based in each case
  on Eq.\ (\ref{Eq_X}) and the respective data sets $n=\{4,6,8\}$
  for $i=1$, $n=\{6,8,10\}$ for $i=2$ and $n=\{2,6,10\}$ for $i=3$ (the latter two in the case $s=\frac{1}{2}$
  only).  The classical $(s \to \infty)$ results from Eqs.\ (\ref{chi_neel_classical}) and (\ref{chi_stripe_classical}) are also shown.}
\label{Ext-M-Field}
\end{figure}

Finally, in Fig.\ \ref{Ext-M-Field}, we show corresponding results for
the zero-field transverse magnetic susceptibility $\chi$ to those
shown in Fig.\ \ref{sStiff} for $\rho_{s}$.
Two things are immediately apparent from Fig.\ \ref{Ext-M-Field}.
Firstly, it is evident that the effect of quantum fluctuations is to
reduce $\chi$ in both quasiclassical AFM phases from its classical
value.  This reduction is greater for the spin-$\frac{1}{2}$ case than
for the spin-1 case, as expected, but is considerable in both.
Secondly, and more importantly, we observe from both Figs.\
\ref{Ext-M-Field}(a) and \ref{Ext-M-Field}(b) a marked tendency for
$\chi$ to vanish at the two QCPs in each case.  This is in marked
contrast to the classical case where $\chi_{{\rm cl}}$, also shown in
Fig.\ \ref{Ext-M-Field}, takes a non-vanishing value $(= \frac{1}{9})$
at the corresponding single classical phase transition at
$\kappa_{{\rm cl}}=\frac{1}{2}$.  Thus, in the classical case, when
the spins are placed in a transverse magnetic field (of magnitude $h$)
it is always energetically favourable to cant the spins, even in the
limit $h \to 0$, and hence $\chi \equiv \chi(h=0)\neq 0$.  By
contrast, in the case of a system of quantum spins, if the lowest
available excited state has a nonzero excitation energy (i.e.,
$\Delta \neq 0$), no excitations are possible (at $T =0$) as
$h \to 0$.  Hence, a positive signature of a spin gap opening up at
any QCP at which quasiclassical magnetic LRO of any sort melts is the
vanishing of $\chi$ there
\cite{Mila:2000_M-Xcpty_spinGap,Bernu:2015_M-Xcpty_spinGap}, indepdendent of the order of the transition.

Turning first to the spin-$\frac{1}{2}$ case, shown in Fig.\ \ref{Ext-M-Field}(a),
the results for $\chi$ show very similar effects (and explanations
thereof) to those seen above in Fig.\ \ref{sStiff}(a) for $\rho_{s}$.
Thus, the shapes of the N\'{e}el and striped curves again reflect the
different orders of their two transitions into the intermediate phase.
Also, on the N\'{e}el side, the three extrapolations agree quite well
with each other except near the QCP at $\kappa_{c_{1}}$.  In
particular, only for the LSUB$\infty(2)$ extrapolation, which employs
solely the highest-order approximants available (i.e., those with
$n=6,8,10$) does $\chi$ vanish.  The corresponding value
$\kappa \approx 0.469$ at which $\chi$ vanishes is again in complete
agreement with the best CCM estimate for $\kappa_{c_{1}}$ (viz.,
$\kappa_{c_{1}} \approx 0.45 \pm 0.02$).  Similarly to the results for
$\rho_{s}$ in Fig.\ \ref{sStiff}(a), those in Fig.\
\ref{Ext-M-Field}(a) for $\chi$ in the striped phase also show a marked
$(4m-2)/4m$ staggering effect, which is again reflected in the three
shown LSUB$\infty(i)$ extrapolants.  Corresponding points at which
$\chi \to 0$ in the striped phase are, for example,
$\kappa \approx 0.583$ for LSUB$\infty(2)$ and $\kappa \approx 0.572$
for LSUB$\infty(3)$, both in reasonable agreement with the best CCM
estimate for $\kappa_{c_{2}}$ (viz.,
$\kappa_{c_{2}} = 0.60 \pm 0.02$).  Thus, the results for $\chi$ for
the spin-$\frac{1}{2}$ case, like those for $\rho_{s}$, reinforce our
earlier conclusions about its $T=0$ phase diagram, while again not
adding significantly to their accuracy for similar reasons.  More
importantly, however, they add considerable weight to our finding that
the intermediate paramagnetic state is gapped over most (or all) of
the region $\kappa_{c_{1}} < \kappa < \kappa_{c_{2}}$, particularly at
and near the QCP at $\kappa_{c_{2}}$.

Finally, turning to the spin-1 case, shown in Fig.\
\ref{Ext-M-Field}(b), we again only have SUB$n$--$n$ results for
values of the truncation index $n \leq 8$, and hence we can only
display the SUB$\infty$--$\infty(1)$ extrapolant.  Again, this has
similar shortcomings to those discussed in relation to $\rho_{s}$ for
$s=1$.  While the extrapolated value for $\chi$ does not now vanish
for either the N\'{e}el or striped phase, the tendency to do so is
clear.  Nevertheless, while these results for $\chi$ certainly do not
contradict those in Fig.\ \ref{Egap_fig}(b) for $\Delta$ for this
case, the evidence for the intermediate state for the spin-1 model to
be gapped remains weak at best, except possibly near the QCP at
$\kappa_{c_{2}}$.


\section{Discussion and conclusions}
\label{conclusions_sec}
In this paper we have implemented the CCM to very high orders of
approximation in order to investigate the frustrated spin-$s$
$J_{1}$--$J_{2}$--$J_{3}$ Heisenberg antiferromagnet on the honeycomb
monolayer lattice.  We have concentrated on the case
$J_{3}=J_{2}\equiv \kappa J_{1}$ and have investigated the $T=0$
quantum phase diagrams of the model for various values of the spin
quantum number $s$.  In particular, we have investigated the phase
structure of the model in the most interesting window
$0 \leq \kappa \leq 1$ of the frustration parameter $\kappa$.  This
includes the classical tricritical point at
$\kappa_{{\rm cl}}=\frac{1}{2}$ of the full $J_{1}$--$J_{2}$--$J_{3}$
model, which is the point of maximum frustration.  The classical model
is the limiting case $s \to \infty$, and along the line
$J_{3}=J_{2}\equiv\kappa J_{1}$ under study the system undergoes a
single GS phase transition at $T=0$, such that for
$\kappa < \frac{1}{2}$ the stable phase has AFM N\'{e}el LRO, and for
$\kappa > \frac{1}{2}$ there exists an IDF of non-planar spin
configurations, all degenerate in energy, which includes the collinear
state with AFM striped LRO as a special case.  Both thermal and
leading-order quantum fluctuations at $O(1/s)$ lift the degeneracy in
favour of the striped state for $\kappa > \frac{1}{2}$.  Our interest
here has been to study whether and how quantum fluctuations change the
structure of the classical phase diagram for various values of the
spin quantum number $s$.

In order to do so we have first calculated very accurate values for
the magnetic order parameter $M$ of both quasiclassical AFM phases for
a variety of values of the spin quantum number $s$.  Clear evidence
thereby emerged that for the two values $s=\frac{1}{2}$ and $s=1$ the
single classical critical point at $\kappa_{{\rm cl}}=\frac{1}{2}$ is
split into two QCPs, $\kappa_{c_{1}} < \kappa_{{\rm cl}}$ and
$\kappa_{c_{2}} > \kappa_{{\rm cl}}$, such that the system maintains
N\'{e}el magnetic LRO for $\kappa < \kappa_{c_{1}}$ and striped
magnetic LRO for $\kappa > \kappa_{c_{2}}$, while the stable GS phase
is a paramagnet, with no discernible magnetic LRO, in the intermediate
regime $\kappa_{c_{1}} < \kappa < \kappa_{c_{2}}$.  Our best estimates
for the QCPs are $\kappa_{c_{1}}=0.45 \pm 0.02$ and
$\kappa_{c_{2}}=0.60 \pm 0.02$ for the spin-$\frac{1}{2}$ model, and
$\kappa_{c_{1}}=0.485 \pm 0.005$ and $\kappa_{c_{2}}=0.528 \pm 0.005$
for the spin-1 model.  By contrast, for all values
$s \geq \frac{3}{2}$, the results for $M$ clearly indicate a direct
first-order transition between the collinear N\'{e}el and striped AFM
phases, at a value $\kappa_{c}(s)$ of the frustration parameter that
depends on $s$ and which appears to approach the classical value
$\kappa_{c}(\infty) \equiv \kappa_{{\rm cl}} = \frac{1}{2}$
monotonically from above.  For example, for the spin-$\frac{3}{2}$
model we found $\kappa_{c}(\frac{3}{2}) \approx 0.517$.

Those findings from the behaviour of $M=M(\kappa)$ were reinforced
from further high-order CCM calculations of the triplet spin gap
$\Delta$ for the three cases $s=\frac{1}{2},1,\frac{3}{2}$.  In
particular, for the spin-$\frac{3}{2}$ model, our results showed that
$\Delta=0$, within very small numerical errors associated with the
extrapolations, over the entire range $0 \leq \kappa \leq 1$,
compatible with the existence therein everywhere of states with
magnetic LRO, whose lowest-lying magnon excitations are gapless.  By
contrast, for the spin-$\frac{1}{2}$ model, clear evidence emerged
that $\Delta \neq 0$ over at least a large part (if not all) of the
intermediate regime $\kappa_{c_{1}} < \kappa < \kappa_{c_{2}}$
(especially near the second QCP at $\kappa_{c_{2}}$), fully compatible
with earlier CCM finding \cite{DJJF:2011_honeycomb} that the
paramagnetic state in this case has PVBC order.  For the remaining
case, $s=1$, the results from $\Delta$ were far less conclusive.
Nevertheless, on balance, the evidence does point towards a gapped
state opening up at (or very near to) the QCP at $\kappa_{c_{2}}$ at
which striped order melts.

In view of these results we decided finally to calculate the remaining
two low-energy parameters (viz., the spin stiffness $\rho_{s}$ and the
zero-field uniform transverse magnetic susceptibility $\chi$).  We
decided also to focus on the two cases $s=\frac{1}{2},1$ in order to
see if we could glean any further evidence from these parameters about
the $T=0$ quantum phase diagram of the spin-1 model, by comparing
their behaviours as functions of $\kappa$ with those of their
spin-$\frac{1}{2}$ counterpart.  For both cases $s=\frac{1}{2}$ and
$s=1$ the results and conclusions from the calculations of $\rho_{s}$
largely supported those from the corresponding calculations of $M$.
In particular, from the behaviour of the two curves $M=M(\kappa)$ and
$\rho_{s}=\rho_{s}(\kappa)$ for both magnetically ordered phases, the
evidence is that for both the spin-$\frac{1}{2}$ and the spin-1 model
the transition at $\kappa_{c_{1}}$ is continuous, while that at
$\kappa_{c_{2}}$ is first-order.

In principle, calculations of $\chi$ can be particularly illuminating
since they can provide direct evidence of a gapped state opening,
viz., at points where $\chi \to 0$.  In the case of the
spin-$\frac{1}{2}$ model the evidence was rather compelling that at
both QCPs, $\kappa_{c_{1}}$ and $\kappa_{c_{2}}$, $\chi$ vanishes,
thereby reinforcing the belief that the intermediate state in this
case is gapped everywhere, compatible with it having PVBC order over
the whole intermediate interval
$\kappa_{c_{1}} < \kappa < \kappa_{c_{2}}$.  By contrast, for the
spin-1 model the evidence was less conclusive, largely due
to the fact that the previously known $(4m-2)/4m$ staggering for CCM
SUB$n$--$n$ approximants of physical observables for honeycomb-lattice
systems is clearly present for $\chi$.  Hence, this makes
extrapolations of $\chi$ particularly problematic for the spin-1 model
where SUB$n$--$n$ approximants to $\chi$ are only really practicable
(even with large amounts of supercomputing resources available, as
here) for values $n \leq 8$ of the truncation parameter, in comparison
with $n \leq 10$ for its spin-$\frac{1}{2}$ counterpart.

In summary, the $T=0$ quantum phase diagrams for the model under study
are now well understood from our CCM calculations for all values of the
spin quantum number, except $s=1$.  The nature of the intermediate
phase for the spin-1 model remains elusive, even after such an
exhaustive study as has been performed here.  While we have obtained
weak evidence that a gapped state opens up as striped order vanishes
below $\kappa_{c_{2}}$, this is far from being compelling.
Furthermore, there is no direct evidence at all for a gapped state
opening above $\kappa_{c_{1}}$ where N\'{e}el order melts.  It would
clearly be of great interest to perform SUB10--10 calculations for $\chi$
in both magnetic phases for the case $s=1$, although it seems
infeasible that these will become available in the near future.  It
might also be of interest to use other alternative techniques for the
spin-1 model, although they will clearly require very high accuracy to
be able to give solid conclusions, on the evidence of the present
work.

\section*{Acknowledgments} 
We thank the University of Minnesota Supercomputing Institute for the
grant of supercomputing facilities, on which the work reported here
was performed.  One of us (RFB) gratefully
acknowledges Leverhulme Trust for the award of an
Emeritus Fellowship (EM-2015-007).  

\section*{References} 

\bibliographystyle{iopart-num}
\bibliography{bib_general}

\end{document}